\documentclass[aps,prd, twocolumn,floatfix,preprintnumbers,superscriptaddress,showpacs]{revtex4-1}
\usepackage{fontenc}
\usepackage{graphicx}
\usepackage{amsmath,amsfonts}
\usepackage{amssymb}
\usepackage{slashed}
\usepackage{color}

\begin{document}

\preprint{JLAB-THY-14-1834}

\title{Analytic Evolution of Singular  Distribution Amplitudes in QCD}

\author{A. V. Radyushkin}

\affiliation{Physics Department, Old Dominion University, Norfolk,
             VA 23529, USA}
\affiliation{Thomas Jefferson National Accelerator Facility,
              Newport News, VA 23606, USA
}
\affiliation{Bogoliubov Laboratory of Theoretical Physics, JINR, Dubna, Russian
             Federation}

        \author{A. Tandogan}  

\affiliation{Physics Department, Old Dominion University, Norfolk,
             VA 23529, USA}
\affiliation{Thomas Jefferson National Accelerator Facility,
              Newport News, VA 23606, USA
}

\begin{abstract}

We describe a    method of analytic  evolution 
 of distribution amplitudes  (DA)
 that have singularities, such as non-zero values at the end-points
 of the support region, jumps at some points  inside 
the support region  and cusps.
We   illustrate  the method by applying it to the 
evolution of a  flat (constant) DA,  antisymmetric 
flat DA and then  use it for evolution of the 
 two-photon generalized distribution amplitude. 
 Our approach has advantages over 
the standard method of expansion in Gegenbauer polynomials,
  which requires infinite number of terms in order to accurately reproduce 
  functions in the vicinity of singular points,   and over a 
  straightforward iteration of  an initial distribution 
  with evolution kernel. The latter  produces logarithmically 
  divergent terms at each iteration, while 
in  our method  the logarithmic singularities are summed 
 from the start, which  immediately produces a 
   continuous curve, with 
only one or two iterations   needed afterwards  in order to get rather 
precise results.

\end{abstract}

\maketitle

\section{Introduction}

Construction of  theoretical   models for Generalized Parton Distributions
(GPDs)  \cite{Mueller:1998fv,Ji:1996ek,Radyushkin:1996nd,Collins:1996fb} 
is an inherent part of their studies. 
These models should   satisfy several nontrivial requirements  
that follow from  most  general principles  of  quantum  field  theory.
In this context, one could mention 
polynomiality \cite{Ji:1998pc},  positivity \cite{Martin:1997wy,Pire:1998nw,Radyushkin:1998es}, 
  hermiticity \cite{Mueller:1998fv},  time reversal invariance \cite{Ji:1998pc},  etc.
These complicated conditions, however, are automatically
satisfied by relevant Feynman diagrams of perturbation theory. 
In particular,  analysis of simple one-loop diagrams 
is the basis of the  factorized DD Ansatz  \cite{Radyushkin:1998es}  
 that is a standard element
of codes generating  models for GPDs. 

In fact, the commonly used version
of  the factorized DD Ansatz involves the assumption about universality of the DD profile function,
which, though supported by one-loop examples, was not shown 
to be a mandatory property of double distributions. 
A possible way to go beyond the one-loop analysis, but still remain  
within the  perturbation theory framework,  is to
incorporate pQCD evolution equations.
Namely, the strategy is to take the expression for some one-loop diagram as the 
starting function for  evolution,  and use evolved patterns 
for modeling GPDs. 

 Implementation of such a program 
faces  some technical  difficulties.
In particular, 
a well known property of GPDs $H(x,\xi)$ is that they are non-analytic 
at border points $x=\pm \xi$.  For one-loop diagrams,
this non-analyticity may take the form of cusps, jumps,
and even delta-functions. Thus, one needs to develop methods 
of evolution for singular initial distributions.  
A simple example of a singular 
distribution is given by  flat  distribution amplitudes $\varphi (x)=$ const   which do not vanish
at the $x=0, x=1$ boundaries of the DA support region.
In our work   \cite{Tandogan:2011cp}, we proposed  a method that
allows to easily establish the major evolution pattern $(x\bar x)^t$ 
for a flat DA, and  provides an algorithm for an analytic calculation
of  corrections to it. The method was also applied to a DA $\varphi (x)={\rm sgn}  (x-1/2)$
that has a jump  at $x=1/2$, in the middle of the support interval,
with DA being antisymmetric with respect to that point.

In the present paper, we  describe  the  method  of analytic evolution 
of singular distribution amplitudes outlined in 
Ref.  \cite{Tandogan:2011cp},  and extend it 
for studying evolution of generalized distribution amplitudes (GDAs)  \cite{Diehl:1998dk}
(our preliminary results on this application have been reported in Ref.  \cite{Tandogan:2014}).
Similarly to GPDs, these functions are non-analytic 
at kinematics-dependent points $x=\zeta, 1-\zeta$ inside the support interval.
The evolution of GPDs is further complicated by the fact that GPD evolution 
kernels also depend on skewness $\zeta$ (or $\xi$). Unlike GPDs, 
  GDAs  evolve according  to  the same $\zeta$-independent  
  Efremov-Radyushkin-Brodsky-Lepage  (ERBL)  evolution equation 
  \cite{Efremov:1979qk,Lepage:1980fj} 
as  usual DAs, which allows us to concentrate  on studying 
implications   due to the non-analytic structure of the 
initial distribution.   A particular object that we consider
is the two-photon   GDA  \cite{Pire:2002ut} related to 
 the reaction $\gamma^{\ast}(q) \gamma(q^{\prime})\rightarrow \gamma(p_1) \gamma(p_2)$.
In the QCD lowest order, it is proportional to the $VV\to VV$ 
ERBL evolution kernel, but the evolution of its $\ln Q^2$ derivative,
in the leading logarithm approximation, is governed by the $qq \to qq$
ERBL kernel, just as in examples considered in   Ref.  \cite{Tandogan:2011cp}.

The paper is organized as follows. In Sec. II, we discuss the basic ideas of our 
method. In particular, we convert the evolution equation to the form in 
which the convolution integral has the structure of  the 
``plus prescription'' with respect to the integration variable $y$. 
The equation is further simplified by choosing the Ansatz that
absorbs the extra term that generates terms logarithmically singular 
at  the end-points of the support region. Applying 
this method  for initially  flat  DA, we find that,  
for whatever small positive value of the evolution parameter $t$, 
the flat DA  evolves into a function vanishing 
 at the end points with its  shape dominated by  the $(x\bar{x})^t$
factor. Then we analytically calculate the lowest corrections 
to this approximation. We also  apply this method to the 
antisymmetric flat DA that initially takes opposite values 
for $x<1/2$ and $x>1/2$. Such a DA is the simplest example
of initial distribution with a jump inside the support region,
in this case in the middle of the region.  For further applications,
we consider the case of 
an antisymmetric    jump 
($\phi (x= \zeta_-) = - \phi (x=\zeta_+)$ 
at an  arbitrary position  $x=\zeta$ inside the support region,
and derive the formulas that are used in Secs. III-V,
where we apply the approach to the evolution 
of the two-photon generalized distribution amplitude $\psi^q (x,\zeta, Q^2)$. 
Its logarithmic derivative with respect 
to $Q^2$ satisfies 
the ERBL evolution equation, with initial 
conditions given by a function $\varphi (x, \zeta) $ that has both jumps 
(discontinuities in the value of the function $\varphi (x, \zeta) $) 
and cusps (discontinuities in the value of the derivative  $\partial \varphi (x, \zeta) /\partial x$))
at the ``border'' points $x=\zeta, x= 1-\zeta$. 
The structure of $\varphi (x, \zeta) $ is discussed in Sec. III, 
where it is proposed to split it into a part that 
has antisymmetric jumps at the border points,
and a continuous remainder that has cusps there. 
Evolution of the ``jump'' part of  two-photon GDA is considered in Sec. IV,
while evolution of the  ``cusp''  part is considered in Sec. V.
Finally, we summarize the paper.

\section{Evolution of Singular  Distribution Amplitudes}

\subsection{Evolution equation: basics}

Our goal is to   study  ERBL evolution of  distribution amplitudes (DAs) 
having   singularities, with particular emphasis on  the analytical
investigation of the behavior of evolved amplitudes in the vicinity 
of singular points. Specifically, we will analyze the evolution governed by
the nonsinglet quark-quark ERBL kernel. 
Particular examples of  DAs whose evolution is described by 
this  kernel include pion DA, $\rho$-meson DA and   (logarithmic derivative of)  
nonsinglet  two-photon  GDAs. In the   leading logarithm  approximation,  
the relevant ERBL 
evolution equation \cite{Efremov:1979qk,Lepage:1980fj} reads 
\begin{align}
  \frac{\partial \varphi (x,t)}{\partial t}=\int_0^1  [V(x,y)]_+\,  \varphi(y,t)dy,
\end{align}
where $t=2C_F \ln\ln(\mu/\Lambda)/b_0$ is the 
leading logarithm 
QCD evolution parameter,  
\begin{align}
V(x,y)=& \left[\frac{x}{y}\left(1+\frac{1}{y-x}\right)\right]\theta(x<y)
\nonumber \\ &
 +
\left[\frac{\bar{x}}{\bar{y}}\left(1+\frac{1}{x-y}\right)\right]\theta(y<x)  
\end{align}
is the  evolution kernel (we use   $\bar {x}=1-x$ and $\bar{y}=1-y$), which 
has singularity  for  $x=y$ regulated by the 
plus  prescription 
\begin{align}
 [V(x,y)]_+=V(x,y)-\delta(y-x)\int_0^1 V(z,y)dz 
\end{align}
with respect to the first argument  $x$.
In explicit form 
 the evolution equation is 
\begin{align}
  \frac{\partial \varphi (x,t)}{\partial t}=
 \int_0^1 \left [ V(x,y) \varphi(y,t)- V(y,x) \, \varphi(x,t) \right ]\, dy 
 \label{evol_expl} \ .
\end{align}

The usual approach \cite{Efremov:1979qk,Lepage:1980fj}  to solve ERBL equation is to expand initial 
DA over  the  eigenfunctions $x(1-x)C_n^{3/2} (2x-1)$
of the evolution kernel.  Each    Gegenbauer projection  then changes 
as a power $t^{-\lambda_n }$  of the evolution parameter. 
 All anomalous dimensions $\lambda_n$ are positive,
except for $\lambda_0$ which is zero,  hence only the $\sim x(1-x)$ part survives
for $t\to \infty$.
 For a pion DA  given by a sum of a few Gegenbauer polynomials,
this method gives a  convenient analytic expression for the DA evolution. 
However,   if the initial DA does not vanish   at the end points,
or has jumps inside the support region,  one should formally take an infinite number
of Gegenbauer polynomials. In practice, this means that 
 one should  sum over a   large 
number of terms to get a reasonably precise  (point by point) 
result for the evolved DA.

Another way is to solve the evolution equation by iterations, 
i.e. write the solution  as a series expansion 
\begin{align}
 \varphi(x,t)=\sum_{n=0}^{\infty} \frac{t^n}{n!} \, \varphi_n(x) \ , 
\label{expansion}
\end{align}
with the functions $\varphi_n(x) $  satisfying the recurrence  relation
\begin{align}
  \varphi_{n+1}(x)
 =  \int_0^1 \left [ V(x,y) \varphi_n(y)- V(y,x) \, \varphi_n(x) \right ]\, dy  \  .
\end{align}

\subsection{Evolution of flat DA}

Taking flat DA $\varphi^F_0(x)=1$ as the simplest example of the function that
does not vanish at the end points, one finds  that the first iteration  
\begin{align}
\varphi^F_1 (x)  =& \int_0^1[V(x,y)-V(y,x)]dy 
 \nonumber  \\ & = 3/2+x\ln\bar{x}+\bar{x}\ln x \equiv v(x)
   \label{phi1} 
\end{align}
has logarithmic singularities  $\ln x$ and $\ln\bar{x} $  at the end points.  
In fact, they just reflect   the $t$-expansion of   a regular  function 
 $(x \bar x)^t$ 
that appears as the all-order summation result, and 
has a  power behavior at the end points. It is easy to obtain that 
the singular, logarithmic  part of $\varphi_1^F (x)$ comes 
from the singular $1/(x-y)$ part 
of the QCD evolution  kernel, which 
contributes 
$  2+\ln(x\bar{x}) $ into the iteration result.

The  singularities  
of $V(x,y)$ and $V(y,x)$  at $x=y$ 
cancel each other  in the   integrand  of
the  evolution equation (\ref{evol_expl}),   even though 
the subtraction procedure in this integral  does not look like  a  ``+''-prescription with respect to 
the integration  variable ``$y$''.  But one can rewrite (\ref{evol_expl}) as 
\begin{align}
 \frac{\partial \varphi (x,t)}{\partial t}= &
\int_0^1 V(x,y)[\varphi(y,t)-\varphi(x,t)]dy
\nonumber \\ &
+\varphi(x,t)\int_0^1[V(x,y)-V(y,x)]dy  \  , 
\label{evol_yplus}
\end{align}
where the first  term  has the desired structure of the ``+''-prescription with respect to 
$y$. The integral in the second term coincides with 
the first iteration of the evolution kernel  with the flat DA 
denoted in (\ref{phi1})   by $v(x)$,
i.e., we can write
\begin{align}
 \frac{\partial \varphi (x,t)}{\partial t}= &
\int_0^1 V(x,y)[\varphi(y,t)-\varphi(x,t)]dy
\nonumber \\ &
+ v(x) \, \varphi(x,t)   \  .
\label{evol_yplus2}
\end{align}
The evolution equation  (\ref{evol_yplus2}) 
can be simplified through 
taking   the Ansatz 
\begin{align}
 \varphi(x,t) = e^{t v(x)}\Phi(x,t)\ ,
\label{ansatz}
\end{align}
resulting in  
\begin{align}
 \frac{\partial{\Phi  (x,t) }}{\partial t} =\int_0^1 V(x,y) \, \left [e^{t [v(y)  -v(x)]}\Phi(y,t)-\Phi(x,t) \right ]\, dy\, .
\end{align}
Expanding  
\begin{align}
\Phi(x,t)=\sum_{n=0}^{\infty} \frac{t^n}{n!} \Phi_n(x) 
\label{expansion_v}
\end{align}
gives   the recurrence  relations
\begin{widetext}
\begin{align}
  \Phi_{n+1}(x)= \int_0^1  V(x,y) 
  & \Biggl \{\sum_{l=0}^n   \frac{n!\, \Phi_l(y) }{(n-l)!\,l!} 
 \left[    v(y) - v(x) \right]^{n-l} 
 -\Phi_n(x)\Biggr \} dy \  .
\end{align}

Keeping only the singular part of the evolution kernel,
i.e., using $v^{\rm sing} (x) = 2+\ln(x\bar{x})$, we obtain   for the first terms,  
\begin{align}
\Phi_1^{\rm sing} & (x) =  \int_0^1 V(x,y)[\Phi_0(y)-\Phi_0(x)]dy \, ,    
\nonumber   \\ 
\Phi_2^{\rm sing}  &(x) =   \int_0^1V(x,y) \bigl [\Phi_0(y)\ln\left( \frac{y\bar{y}}{x\bar{x}}\right)
 +\Phi_1(y )-\Phi_1(x)\bigr ]dy  \,  , 
\nonumber   \\ 
 \Phi_3^{\rm sing}  & (x)   =  \int_0^1V(x,y)\left  [\Phi_0(y)\ln^2\left(\frac{y\bar{y}}{x\bar{x}}\right)
 +2\Phi_1(y)\ln \left(\frac{y\bar{y}}{x\bar{x}}\right) 
 +\Phi_2(y)-\Phi_2(x)   \right ]dy  \,   . \ \   \
\end{align}

For the flat initial DA, i.e., $\Phi(x,0)=1$,  the first correction $\Phi_1 (x)$ vanishes,
$\Phi_1(x,0)=0$, and then  
\begin{align}
\Phi_2^{\rm sing}  (x)&= -2\ln x\ln\bar{x}   \ , \\
\Phi_3^{\rm sing}  (x)&=3\ln(x\bar{x})\ln x \ln\bar{x}+2\left [\ln x \, \text{Li}_2(x) 
+\ln\bar{x}\, \text{Li}_2(\bar{x}) \right ]  
-4\, \left [ \text{Li}_3(x)   + \,  \text{Li}_3(\bar{x}) \right ]+8 \zeta(3)  \ .
\end{align}

For  the total kernel, when  $v(x) = 3/2+x\ln\bar{x}+\bar{x}\ln x $, we have 
\begin{align}
 \Phi_2(x)
=&
x\ln x[1+(x-1/2)\ln x]+\bar{x}\ln\bar{x}[1+(x+1/2)\ln\bar{x}]-\ln x\ln\bar{x}
 + x\text{Li}_2\left(- \frac{\bar x}{x}\right)+
\bar{x}\text{Li}_2\left(-\frac{x}{\bar x}\right) 
\end{align}
for the second iteration. The third iteration in this case is given 
by a rather long expression.

\end{widetext}

   \subsection{Normalization}

As far as  $$\int_0^1 [V(x,y)]_+ dx =0\ , $$ evolution does not change the normalization integral
for $\varphi (x,t)$. In particular, if we expand $\varphi (x,t)$  in  $t$
\begin{align}
\varphi (x,t)= \sum_{n=0}^\infty  \varphi_n (x)\,  \frac{t^n }{n!} \ ,
\end{align} 
we should have
\begin{align} 
 \int_0^1  \varphi_n (x) \, dx = \delta_{n0}  \ . 
\end{align} 
However,  when we take the singular kernel case 
  Ansatz 
  $$ \varphi^{\rm sing} (x,t)=e^{t(2+\ln(x\bar{x}))}\Phi(x,t) \ , $$
   the $(t\ln x\bar{x})^N$ terms are summed to all orders, while the series over $\varphi_n(x)$ is restricted to some   finite order 
 $N$. 
As a result, the approximants $\varphi_{(N)}(x,t)$  are 
not normalized to 1.
In particular, if we keep the terms up to $\Phi_2 (x)$, the normalization integral  is  given  by
  \begin{widetext}
\begin{align}
I_2 (t) &  \equiv   \int_0^1 \varphi_{(2)}^{\text{sing}}(x,t) \, dx
=e^{2 t}  \frac { \Gamma^2(1+t))}{  \Gamma(2+2t) } 
   \left\{ 1-t^2 \left[(H_t-H_{1+2 t})^2-\psi_1(2+2 t)\right] 
  \right\} \  \  \ 
\end{align}
with  $H_n$ being  harmonic numbers and $\psi_k$  the  polygamma function.
One can check that
$I_2 (t) =1 + {\cal O} (t^3)$. 
For the next approximation, i.e., for $\varphi_{(3)}^{\text{sing}}(x,t) $,
the normalization integral is \mbox{$I_3 (t) =1 + {\cal O} (t^4)$,} etc.
For   $\varphi_{(N\rightarrow \infty)}(x,t)$, the normalization integral $I_N(t) $ will tend to 1 for all $t$.

The normalization integrals versus $t$ are shown in Fig.\ref{norm}a. 
For approximations involving $\Phi_0 (x)$ and  $\Phi_2 (x)$,  the calculations
were done  analytically,  while the curve  corresponding to inclusion of  $\Phi_{3}^{\text{sing}}(x,t) $
was  calculated numerically. 
As   seen from   Fig.\ref{norm}a, adding more terms brings normalization closer to $1$.

\begin{figure}[h]
\centering
\includegraphics[width=6.25cm]{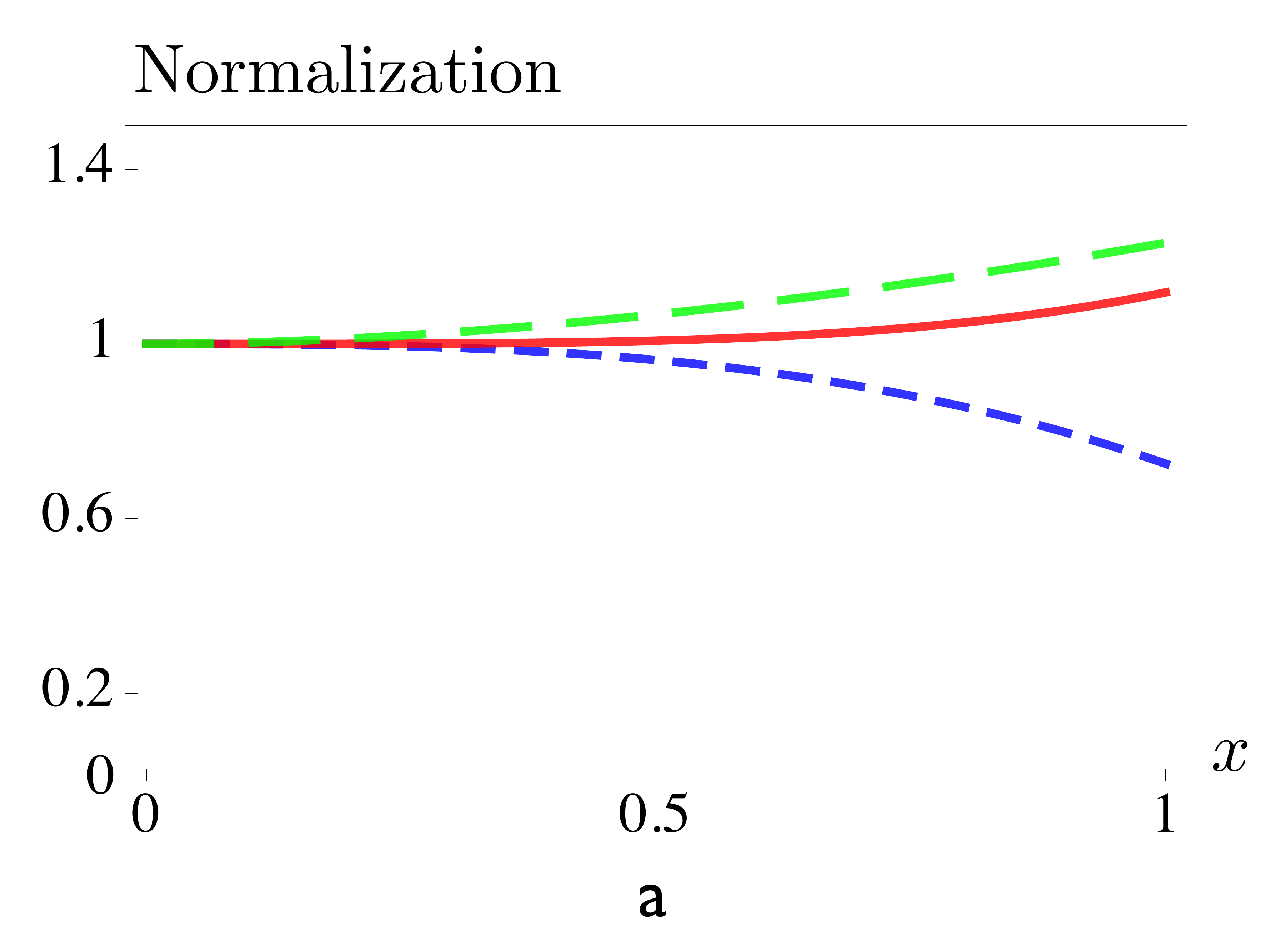}
\includegraphics[width=6.25cm]{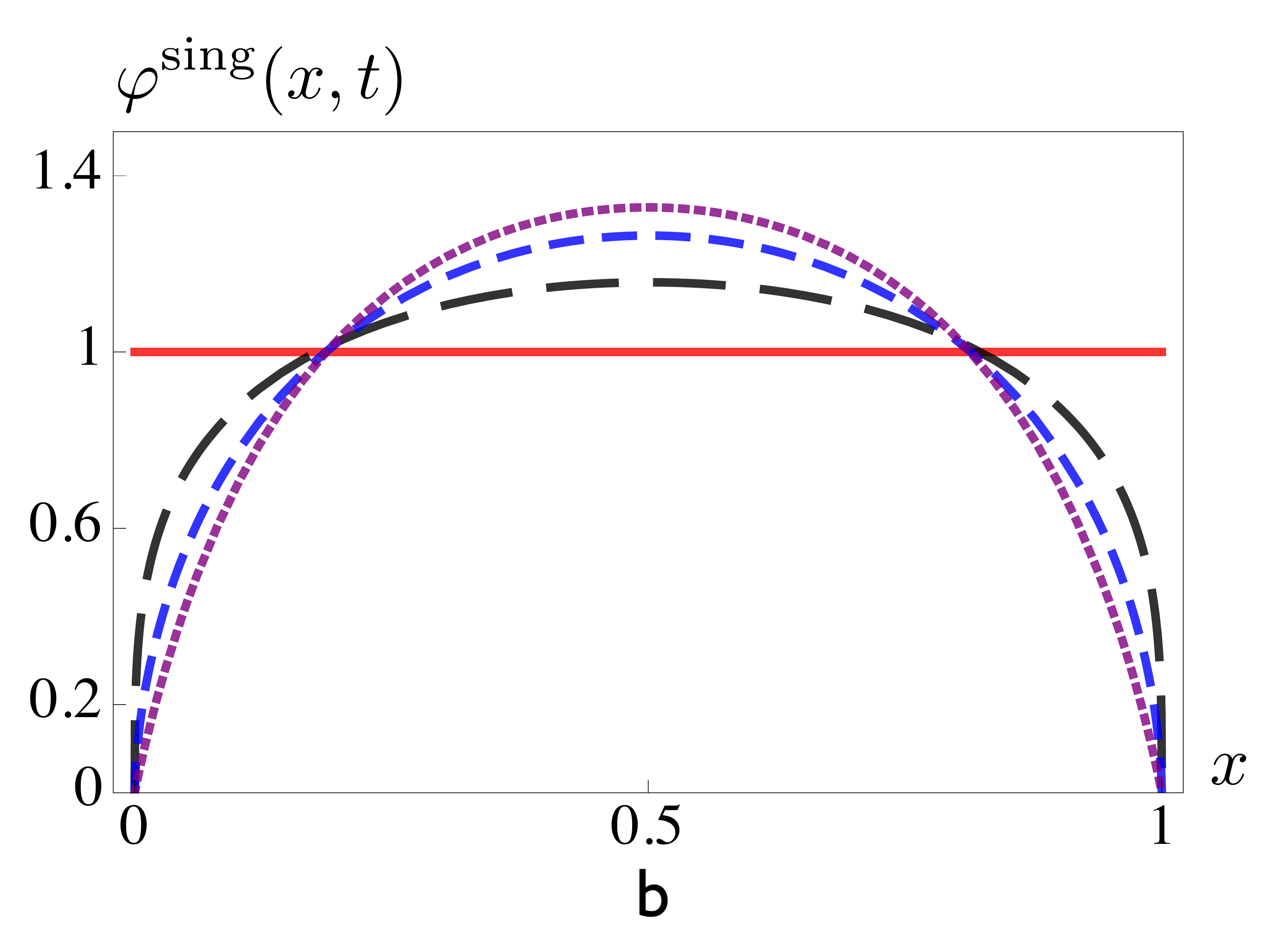}
\caption{ 
{\bf a) } Normalization factor calculated for terms  including only $\Phi_0(x)$ (short-dashed line), 
 $\Phi_0(x)$ and $\Phi_2(x)$ (long-dashed line) and  $\Phi_0(x)$, $\Phi_2(x)$ and $\Phi_3(x)$ 
   (solid line).
{\bf b)}  Evolution of  the flat DA  under the singular part of the evolution kernel: 
the curves shown  correspond to $t=0, t=0.3, t=0.6, t=1.0$.}
   \label{norm}
\end{figure}

In this  situation, it makes sense to introduce the ``normalized Ansatz'', in which $\varphi (x,t)$ 
is approximated by the ratio 
$$\nu_N (x, t) \equiv \varphi_{(N)}(x,t)/I_N(t) \ , $$
 so that 
the correct normalization of the $N$th approximant is guaranteed for all $t$.
In particular, this gives
\begin{align}
 \nu_2 (x, t) =&   (x\bar{x})^t \, \frac{\Gamma(2+2t) }{\Gamma^2(1+t)} 
\frac{1 -t^2 \ln x \, \ln \bar x   }{ 1-t^2 \left[(H_t-H_{1+2 t})^2-\psi_1(2+2 t)\right] }  \ .
\end{align}
As seen from this formula (and also from Fig.\ref{norm}b), the initial flat 
 function immediately (for whatever small positive $t$)
evolves into a function vanishing 
 at the end points with its  shape dominated by  the $(x\bar{x})^t$
factor.

\end{widetext}

In case of  the total kernel,   we have
\begin{align}
\varphi_{(2)}  (x,t)=&  e^{3t/2} \, (x^{\bar{x}}\bar x^x)^t  
\left (1+\frac{t^2}{2!}\Phi_2(x)
\right ) \ .
\end{align}
Unfortunately,  for this form it is  impossible to analytically 
calculate the normalization integral even for the lowest term.
Compared  to the Ansatz used for the singular part of the
evolution kernel,    the  Ansatz
\begin{align}
 \varphi(x,t)=e^{t v(x) }  \Phi(x,t)=e^{3t /2 } 
(x^{\bar{x}}\bar x^x)^t \Phi(x,t)
\end{align}
has an extra overall factor 
$
[e^{-1 /2 }x^{-x} \bar x^{-\bar x}]^{t}  \ .
$
Note that the function
$x^{-x} \bar x^{-\bar x} $ 
is finite at the end points $x=0, \, 1$, where it 
takes its minimal value for the interval $[0,1]$
(equal to  1), 
and has a maximum for $x=1/2$, where it  equals 
2. Thus, the factor $
e^{-1 /2 }x^{-x} \bar x^{-\bar x}
$
enhances the $x\bar x$ profile in the middle 
(by $2/\sqrt{e}\approx 1.2$ factor)
and suppresses it at the end points (by $1/\sqrt{e}\approx 0.6$).
This is a rather mild  modification, and what is most important,
it does not change the $\sim x^t$ (or $\sim \bar x^t$) behavior at the end points.
So,  it makes sense to  use the expansion 
\begin{align}
 [x^{-x} \bar x^{-\bar x}]^{t}= \sum_{n=0}^\infty (-1)^n 
(x \ln  x + \bar x \ln \bar x)^n \, \frac{t^n}{n!} \ ,
\end{align}
in powers of $t$ 
and combine it with the expansion  for $\Phi (x,t)$.

\begin{widetext} 

This corresponds to Ansatz
\begin{align}
\varphi (x,t)=&  (x\bar{x})^t \,  e^{3t/2} \,
\left [\tilde \Phi_0(x)+t \,  \tilde \Phi_1(x)+\frac{t^2}{2!}\tilde \Phi_2(x)+ \ldots 
\right ] \ ,
\end{align}
whose  expansion coefficients $\tilde \Phi_n (x)$ can  be straightforwardly 
obtained  from $ \Phi_n (x)$'s calculated using $v^{\rm sing} (x)$. 
In particular, 
\begin{align} 
\tilde \Phi_1 (x) =& -(x \ln  x + \bar x \ln \bar x)\ , \\
\tilde \Phi_2 (x) =&  \Phi_2 (x) +  (x \ln  x + \bar x \ln \bar x)^2\ . 
\end{align}

For the lowest terms, we can analytically  calculate
the normalization integral:
\begin{align}
I_1(t)  \equiv  & \int_0^1 \varphi_{(1)} (x,t) dx=e^{3 t/2} 
\frac{ \Gamma^2(1+t)}{\Gamma(2+2t)}
 \left  (1 -\frac{t}{2(t+1)} + \frac{t}{2}  (-H_t+H_{1+2 t})\right ) \!
  ,
\end{align}
where  $H_n$ are harmonic numbers. 
Fig.\ref{normf}a shows the normalization versus $t$.

Again, we may switch to the normalized Ansatz   formed by the ratio $ \varphi_{(N)} (x,t)/I_N(t)$.
For a flat initial distribution, this gives
\begin{align}
 \nu_1 (x, t) & =  (x\bar{x})^t  \, \frac{\Gamma(2+2t) }{\Gamma^2(1+t)} 
\frac{1 -t \, (x \ln  x + \bar x \ln \bar x)  }{ 1-{t}/{2(t+1)} + {t}(-H_t+H_{1+2 t})/2  }  \ .
\end{align}
The results are illustrated by   Fig.\ref{normf}b. 

\begin{figure}[h]
\centering
\includegraphics[width=6.25cm]{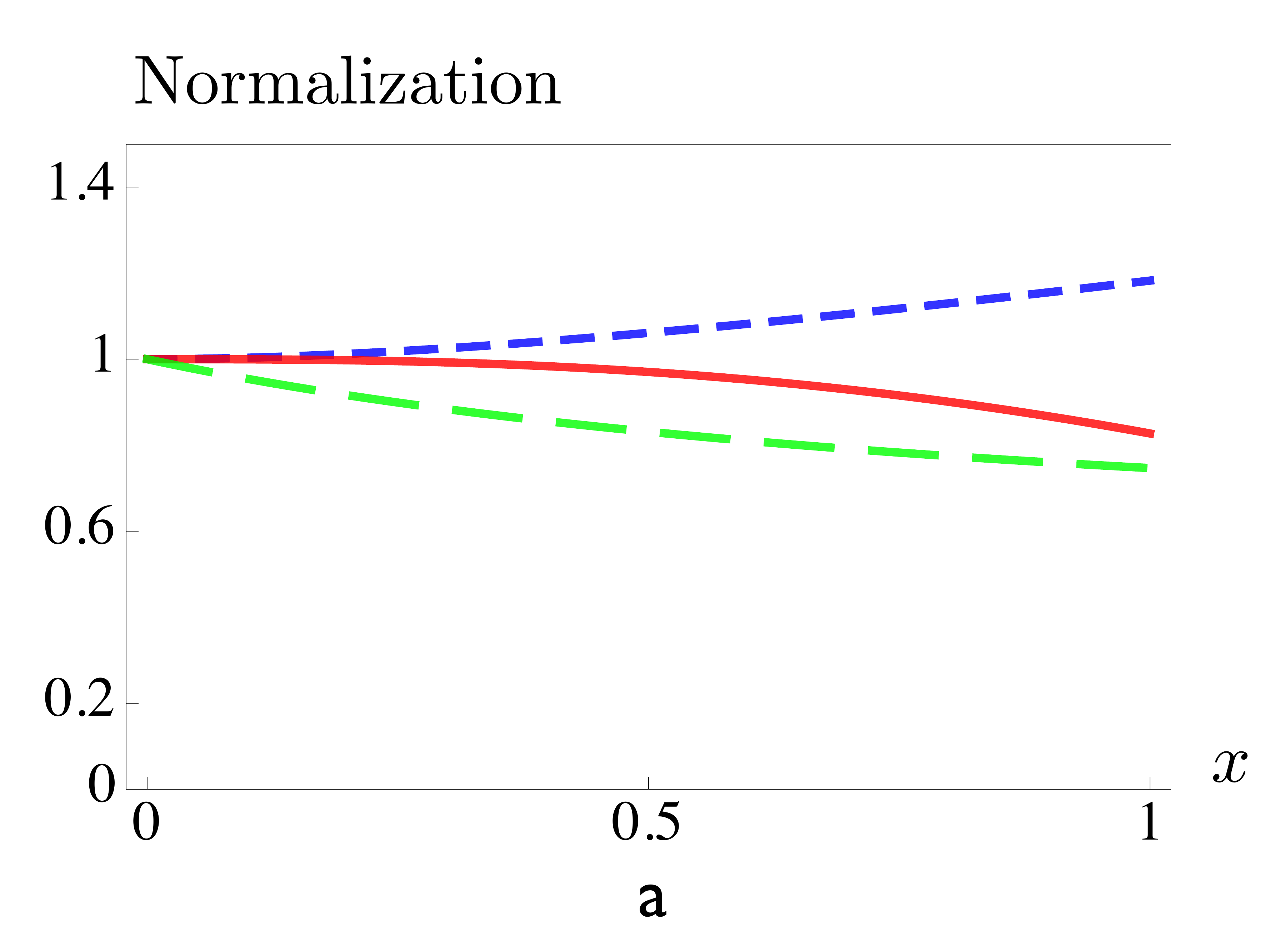}
\includegraphics[width=6.25cm]{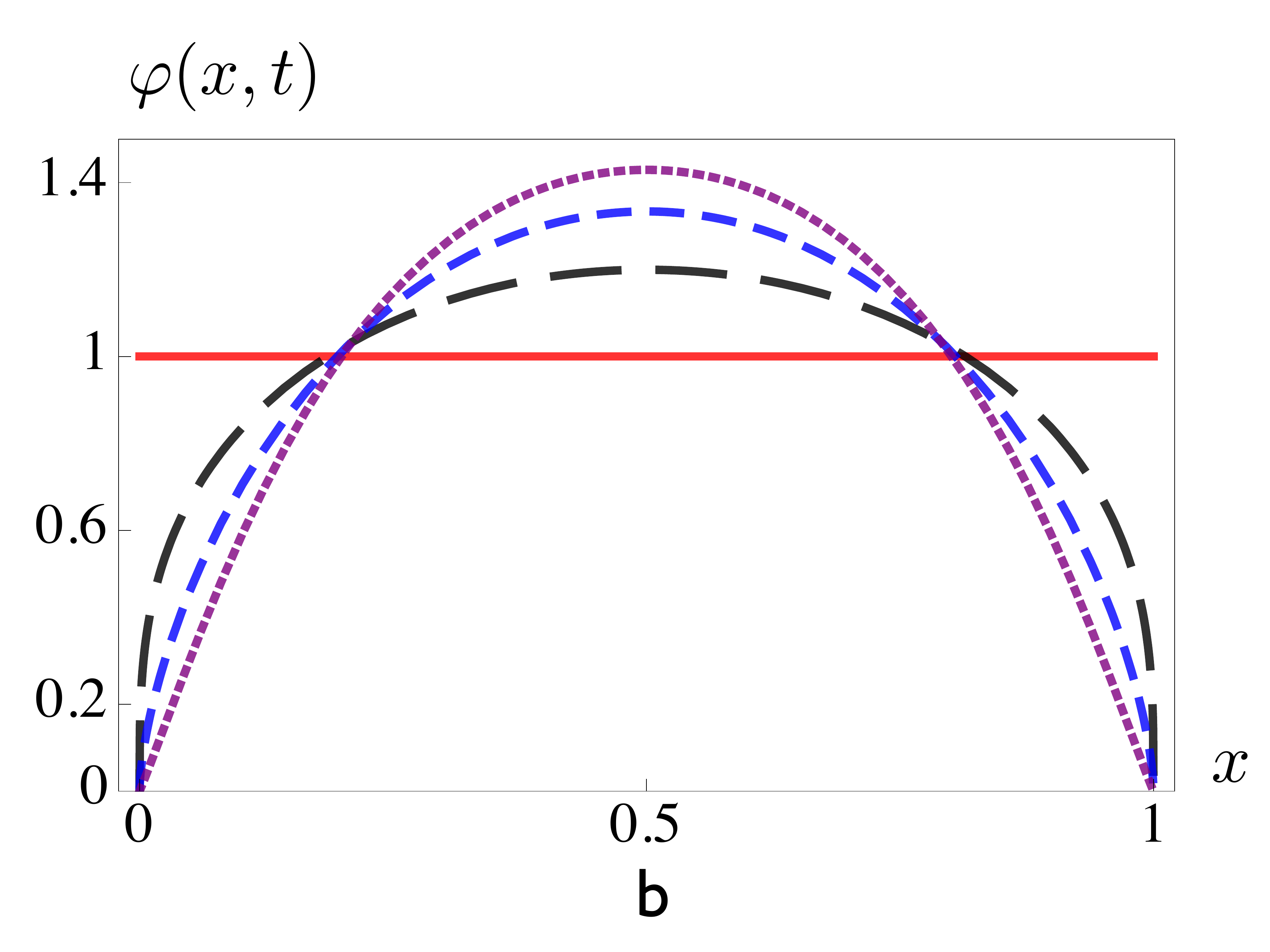}
\caption{ 
 Same as Fig. 1  for the case of   full kernel.
}
   \label{normf}
\end{figure}

\end{widetext} 

\subsection{Evolution of Anti-Symmetric Flat DA}

\subsubsection{Singular Part}

Evolution equations  may be applied also in situations when the distribution amplitude
is  antisymmetric with respect to the change $x \to 1-x$.
An  interesting  example is the $D$-term  function $d(x)$  that appears in 
generalized  parton  distributions. Thus,  let us  consider evolution of the DA  that
initially has the form 
\begin{displaymath}
\varphi_0(x)=\begin{cases} \,\,\,\,1 & 0<x\leqslant 1/2   \ , \\
 -1 & 1/2<x<1  \ .
\end{cases}
\end{displaymath}
The $v(x)$  function 
 is the same, since it depends on the kernel $V(x,y)$ only.
Thus we can use the  Ansatz (\ref{ansatz})  and expansion (\ref{expansion}).
Since $\varphi_0(x)$ is  not just a constant, the first expansion coefficient $\Phi_1 (x)$
is  nonzero. 
Let us  start with the singular part of the kernel.
Then we get
\begin{displaymath}
 \Phi_1(x)=\begin{cases} -2 \ln\left[\frac{\bar{x}}{1-2 x}\right] & 0<x\leqslant 1/2  \ , \\
\,\,\,\,\,2 \ln\left[\frac{x}{-1+2 x}\right] & 1/2<x\leqslant 1  \ .
              \end{cases}
\end{displaymath}
We see that there  are logarithmic terms $ \ln |1-2 x|  $
singular for $x=1/2$. These    terms  are natural, since  each   half 
of the antisymmetric DA on its interval is expected to
 evolve similarly to a  flat DA
on the $0 \leq x\leq1$ interval. 
This observation suggests the Ansatz 
\begin{eqnarray}
 \varphi(x,t)=  e^{2t} (x\bar{x})^t |1-2x|^{2t}\Phi(x,t) \ .
\end{eqnarray}
          \begin{widetext} 
With this definition of $\Phi (x,t)$, the   \mbox{$\ln|1-2x|$}  terms are eliminated from
 $\Phi_1(x)$: 
\begin{eqnarray}
 \Phi_1(x)=-2\ln\bar{x}  \ \theta ( 0<x\leqslant 1/2) - \{x \to \bar x \} \ .
    \end{eqnarray}    
For  the expansion component $\Phi_2(x)$, we have
\begin{align}
 \Phi_2(x)=& \theta ( 0<x\leqslant 1/2)
 \Biggl \{
-\frac{2\pi ^2}{3}+5 \ln^2\bar{x}+\ln(1-2x)[4\ln 2+4\ln x-2\ln\bar{x}+\ln(1-2x)]-2\ln\bar{x}\ln x
\nonumber\\   &
 -4\text{Li}_2(x)+4\text{Li}_2(2x) 
+2\text{Li}_2\left[\frac{x}{\bar{x}}\right]
+2\text{Li}_2\left[\frac{x}{2x-1}\right]+4\text{Li}_2\left[\frac{1-2x}{\bar{x}}\right]
\Biggr \} - \{x \to \bar x \} \   .
\end{align}

 The graphical results for the expansion components are shown in Fig.\ref{anti}a,b).
\begin{figure}[h]
\includegraphics[height=4.15cm]{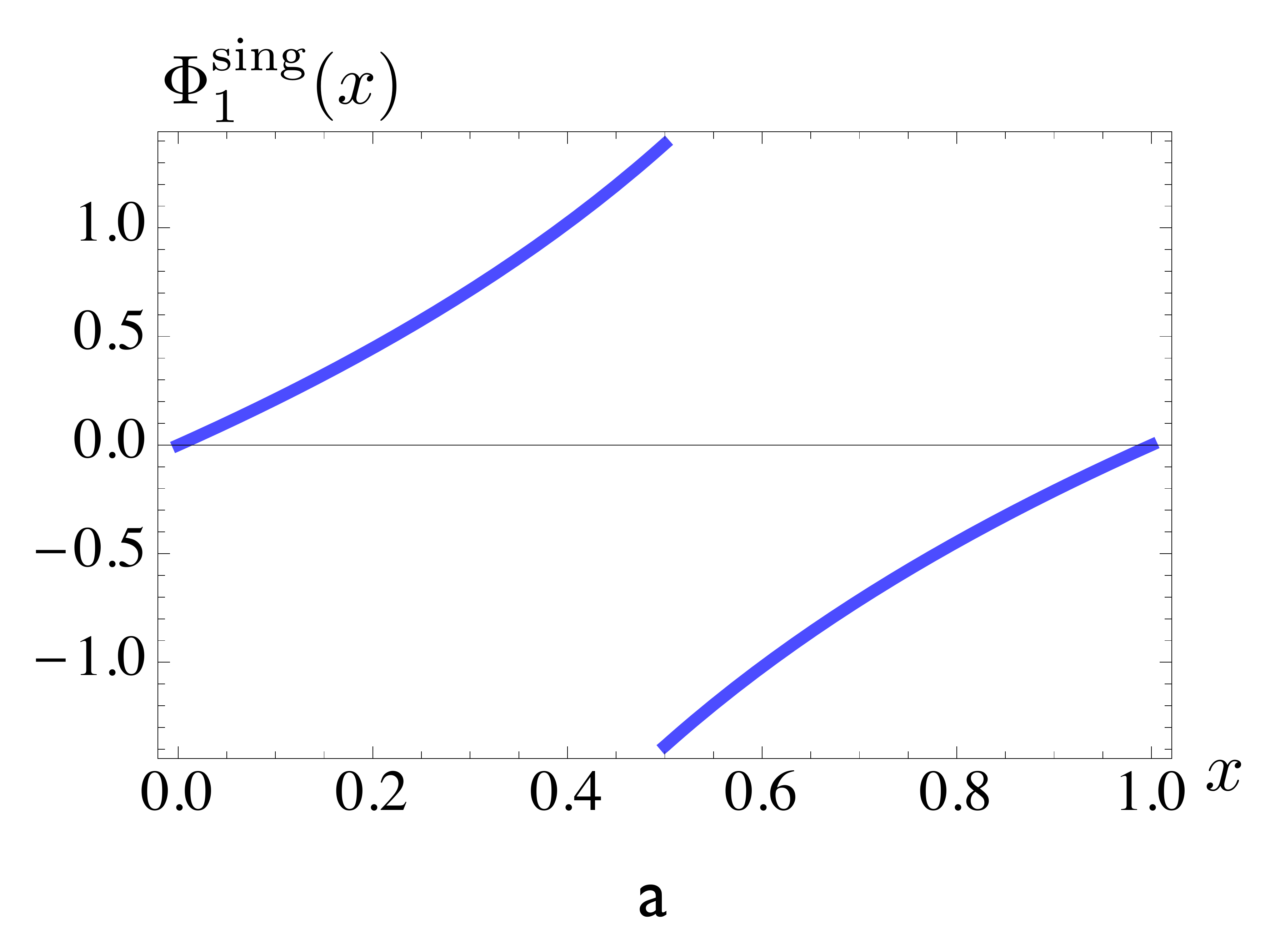}
\includegraphics[height=4.15cm]{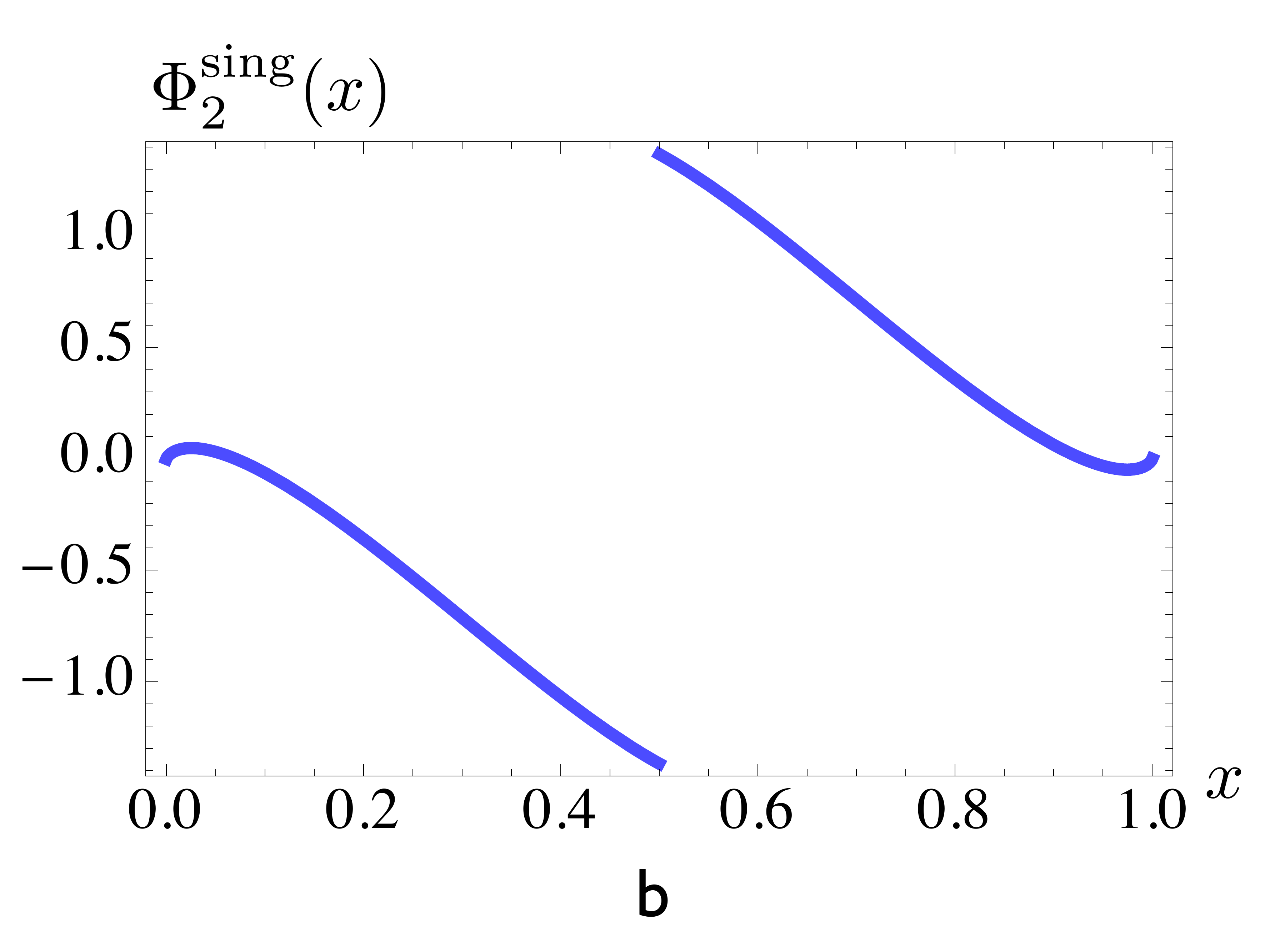}
\includegraphics[height=4.15cm]{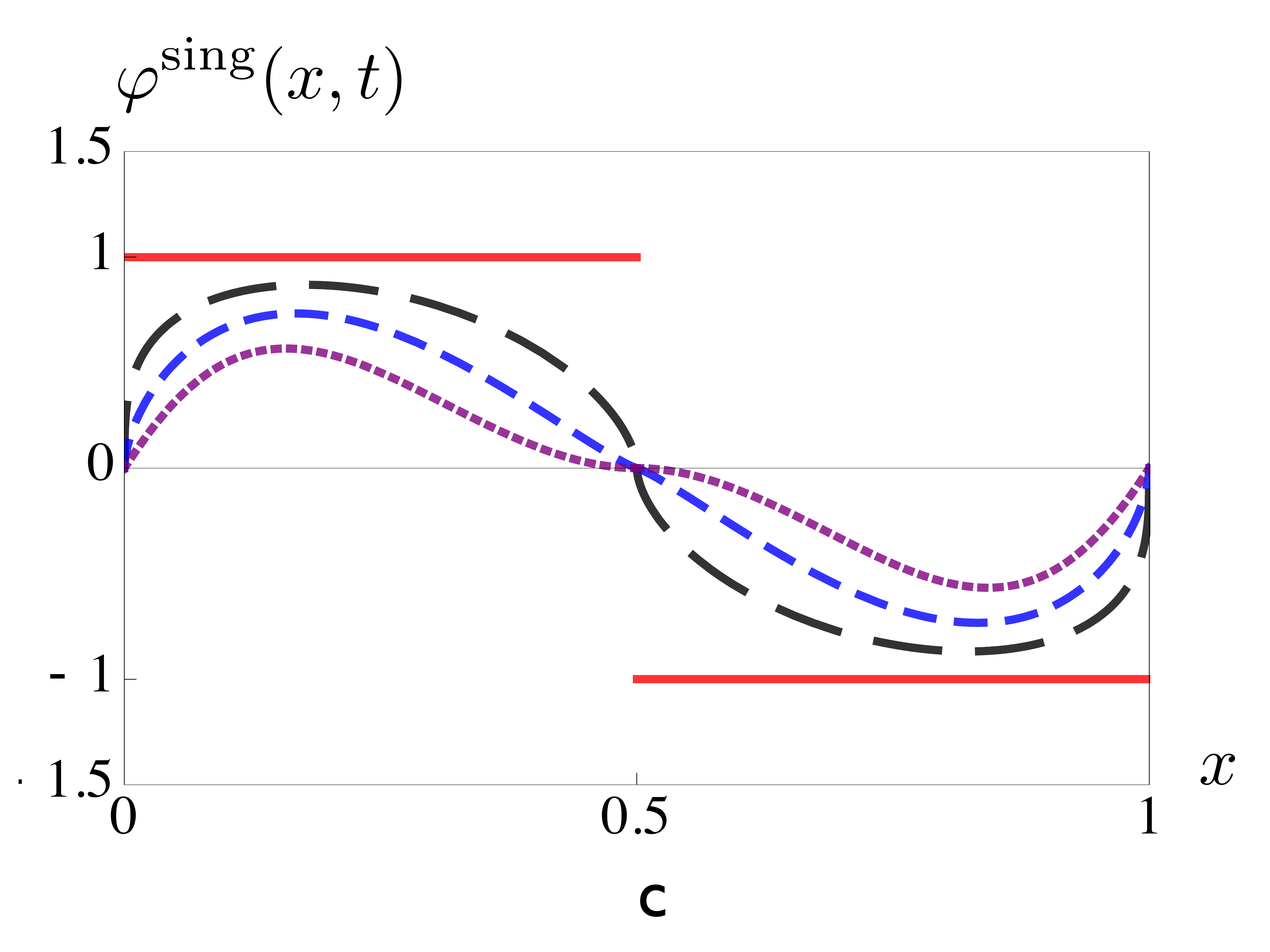}
\caption{Expansion components  $\Phi^{\text{sing}}_1(x)$   (a) and  
  $\Phi^{\text{sing}}_2(x)$ (b)   for the antisymmetric DA in case of singular part of the kernel; 
c) evolution of antisymmetric DA under the singular part of the kernel for  
$t=0, t=0.3, $ $ t=0.6, t=1$.}
\label{anti}
\end{figure}

The evolution of $\varphi_{\text{sing}}(x,t)$ to this accuracy can be obtained from 
\begin{align}
 \varphi(x,t)=& e^{2t} (x\bar{x})^t|1-2x|^{2t}
 \left(\varphi_0(x)+t \Phi_1(x)+\frac{t^2}{2!}\Phi_2(x)\right) \ .
\end{align}

As  shown in Fig.\ref{anti}c, the initial step function evolves into a function which is zero at the end  points
and in   the middle point.

\subsection{Adding  Non-Singular Part of the Kernel}

Since the  nonsingular part  does  not add $\ln x$  and $\ln \bar x$  terms to  $v(x)$, 
we may  proceed with the same Ansatz 
\begin{eqnarray}
 \varphi(x,t)=&e^{3t/2} (x\bar{x})^t |1-2x|^{2t} \Phi(x,t)\, ,
\end{eqnarray}
but the expansion components change (see Fig.\ref{antif}a,b).
\begin{figure}[thb]
\centering
\includegraphics[width=5.5cm]{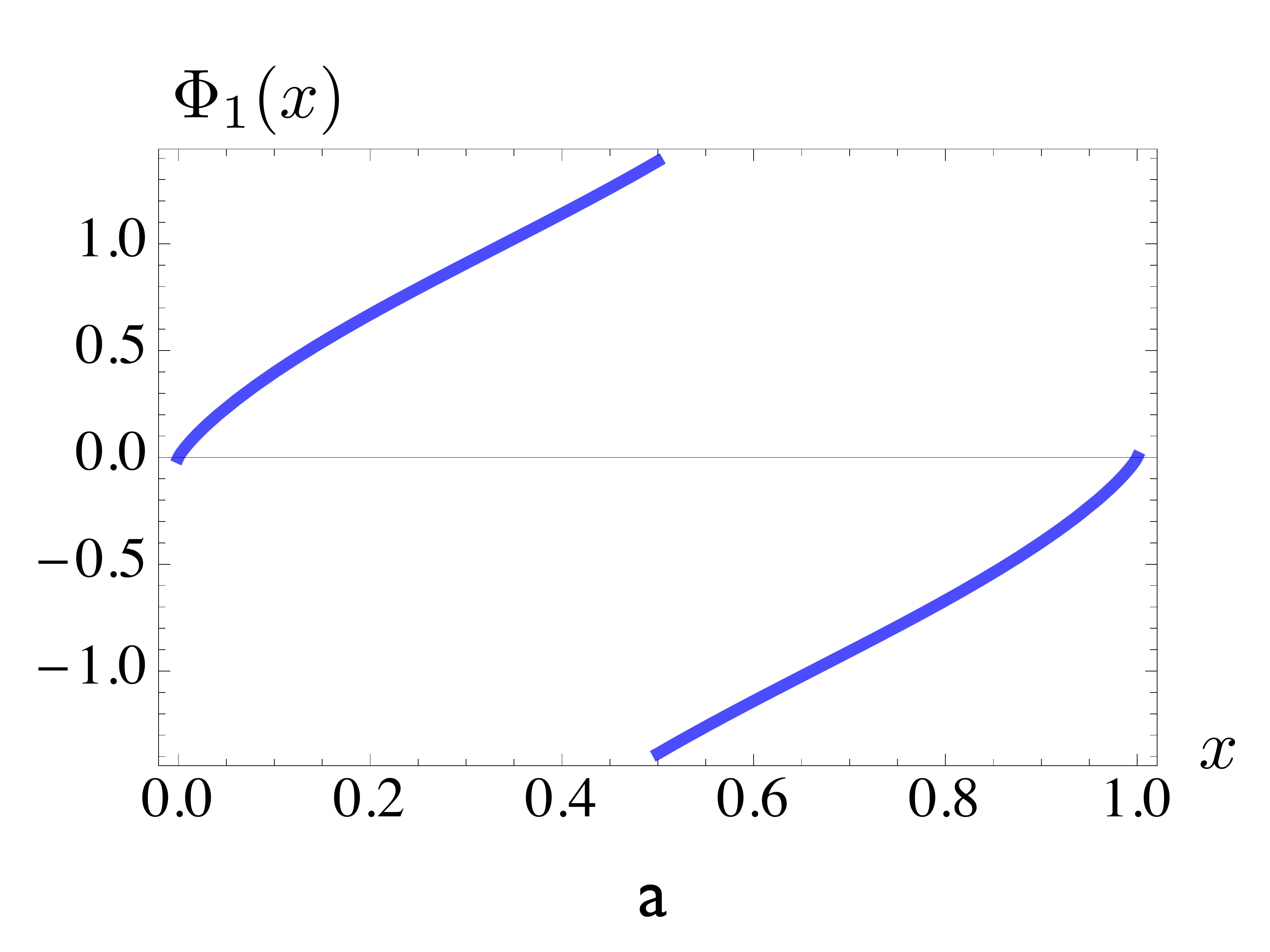}
\includegraphics[width=5.5cm]{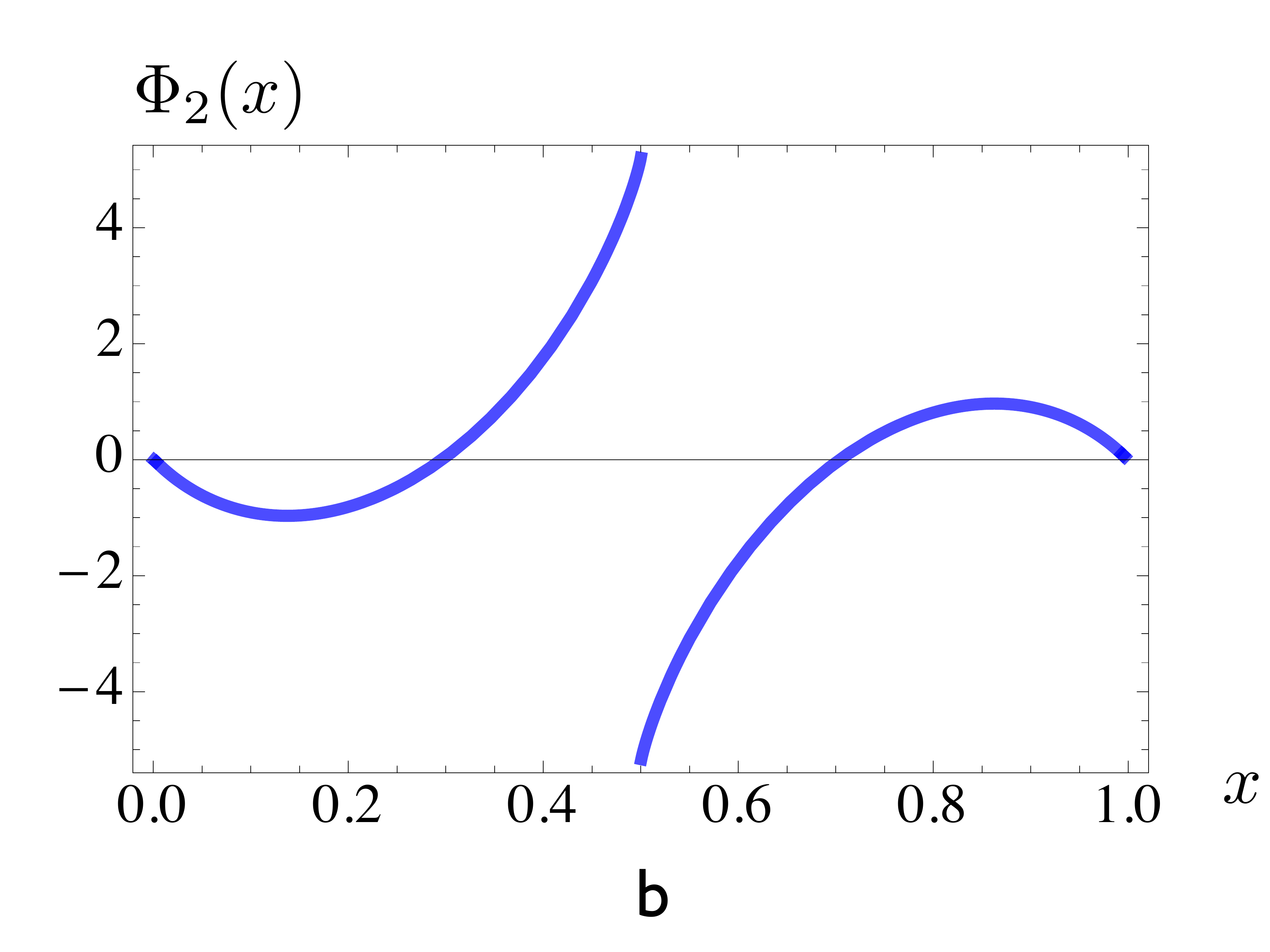}
\includegraphics[width=5.5cm]{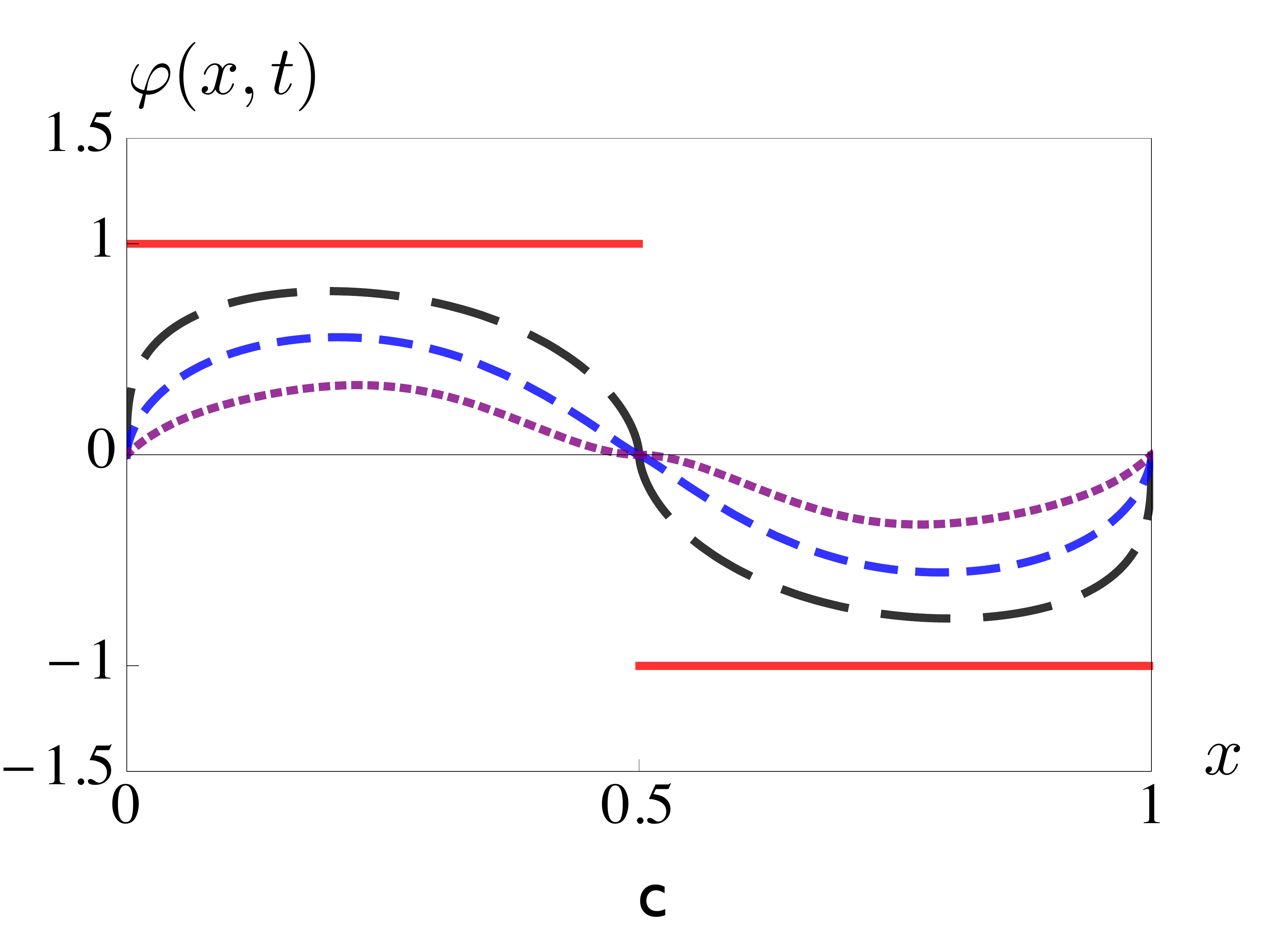}
\caption{ Expansion  components $\Phi_1(x)$  (a) and $\Phi_2(x)$  (b) in the case of the full kernel;  c)
evolution of  the antisymmetric DA in the  full kernel  case for 
 $t=0,$  $ t=0.3, t=0.6,  t=1$.}
\label{antif}
\end{figure}

The distribution amplitude is  now  built using   
\begin{align}
\varphi(x,t)=& e^{3/2t} (x\bar{x})^t|1-2x|^{2t}
\left(\varphi_0(x)+
t \Phi_1(x)+\frac{t^2}{2!}\Phi_2(x)\right),
\end{align}
 For the first coefficient we have 
\begin{eqnarray}
 \Phi_1(x)=\theta ( 0<x\leqslant 1/2) \bigl \{ -2x \ln 2 - 
 2\ln\bar{x} - (\bar{x} \ln\bar{x} + x \ln x) \bigl \}  - \{x \to \bar x \} \ ,  
\end{eqnarray}
and for the second, 
\begin{align}
 \Phi_2(x)=&\theta ( 0<x\leqslant 1/2 )
\Biggl \{ -\frac{\pi ^2}{2}  \bar{x}+2\ln^2+2x \ln 2+\ln\bar{x}[2+8\ln 2+17\ln\bar{x}]-(x-2\ln(1-2x)[\ln\bar{x}-\ln x]
\nonumber\\ &+2(5x-3) \ln 2\ln(1-2x)-\ln^2(1-2x)-6x\ln^2\bar{x}+\ln\bar{x}\ln x+x \ln x(4\ln2+\ln\bar{x}+\ln x\nonumber\\ &+2\ln(1-2x)+2\ln\bar{x})
+2 \bar{x}\bigg[ \text{Li}_2 \left(\frac{1-2x}{2\bar{x}}\right)+\text{Li}_2\left(\frac{1-2 x}{\bar{x}}\right)+\text{Li}_2(x)
-\text{Li}_2(2 x)\bigg] \nonumber\\ &-(1+2 x) \text{Li}_2\left(-\frac{x}{\bar{x}}\right)
 +\text{Li}_2\left(\frac{x^2}{\bar{x}^2}\right)-2 \text{Li}_2\left(\frac{x}{2 x-1}\right)\Biggl \}
 - \{x \to \bar x \} \ . 
\end{align}
As may be seen from Fig.\ref{antif}c, the resulting curves are rather close to those
obtained when only singular part of  the kernel was taken into account.
Thus, we observe that the $t^n$ series   converges rather rapidly as far 
 as $t \lesssim 1$. When $t \gtrsim 1$, the DAs is   close to the asymptotic
 form, and one can switch to the solution in the form of Gegenbauer expansion.  
 
\end{widetext} 

\subsection{Evolution of  jumps}

Another example of singularity is given by  DAs  with  a  jump, the simplest case being 
\begin{displaymath}
\Phi^J_0(x, \zeta; a,b)=\begin{cases} \,\,\,\,a & 0<x\leqslant \zeta   \ , \\
  \,\,\,\, b  & \zeta <x<1  \ .
\end{cases}
\end{displaymath}
The part of the  first iteration  $\Phi_1(x)$ generated by the  
singular part of the kernel 
\begin{displaymath}
 \Phi^{J\, {\rm sing}}_1(x, \zeta; a,b)=\begin{cases} \,\,\,\,\, (a-b) 
 \ln\left[\frac{\zeta-x}{(1- x)\zeta}\right] & 0<x\leqslant \zeta   \ , \\
\,-(a-b) 
 \ln\left[\frac{x-\zeta}{(1- \zeta)x}\right] & \zeta <x\leqslant 1  
              \end{cases}
\end{displaymath}
contains   logarithmic terms $ \ln |\zeta - x|  $
singular for $x=\zeta$.  Their structure may be understood in the following way. 
The original function $\varphi^J_0(x, \zeta; a,b)$
may be represented as a sum of a constant $\frac12 (a+b)$ and 
a function \mbox{$\frac12 (a-b) {\rm sgn}(\zeta -x)$}  that  
 jumps  by $b-a$ at the point $x=\zeta$.
 The constant part has no singularities at $x=\zeta$, so one can apply 
 the original 
 $\frac12 (a+b)(x\bar x)^t\Phi$ Ansatz to it, while for the jumping part 
 one  may use the   Ansatz 
 \begin{align}
 \Phi^J   & (x, \zeta; a,b;t)= \frac{a-b}{2} 
 e^{2t} (x\bar x)^t   \left [ \left (\frac{1-x/\zeta}{1-x}
 \right ) ^{2t} \theta (x <\zeta) \right.  
  \nonumber \\&
   \left. + \left (\frac{
 1-\bar x/\bar \zeta}{1-\bar x} \right )^{2t} \theta (x >\zeta)  \right ] 
 \Phi (x, \zeta) + \Psi (x, \zeta,t) \   .
  \end{align}  
The part containing square brackets is intended to take care of the evolution of the jump at \mbox{$x=\zeta$.}
However,  this part  by construction vanishes at
 $x= \zeta$, while one would expect that evolution tends to convert 
$\varphi^J(x, \zeta; a,b)$  into a universal \mbox{$\zeta$-independent } 
function proportional to  $x \bar x $ or $x \bar x (1-2x)$
(depending on the symmetry of the function).
Thus, there should be also a part regular at the jump point.
 The function  $\Psi (x, \zeta,t) $
is introduced to satisfy  this requirement.
It vanishes for $t=0$, but eventually becomes the dominant 
part.

 Let us discuss a more general case,   when a function 
 has antisymmetric jumps at some locations $x= \zeta_i$.
 ``Antisymmetric'' means that the function approaches 
 opposite values on the sides of a  jump,
 so that  ``on average'' it is zero  at the jump points. 
Then one can 
 try  the Ansatz 
 \begin{align}
  \varphi(x,t) = \Phi(x,t) +  \Psi (x,t) \ ,
 \label{ansatz2}
 \end{align}
 where  
 \begin{align}
   \Phi(x,t)= e^{t [v(x)+ w(x)]}\Phi_0  (x) 
   \  ,
   \end{align}
    with 
  the function $w(x) \sim \sum_i \ln |x-\zeta_i | $  intended to  
 absorb  major features of  the evolution  
of  the starting distribution 
 \mbox{$  \varphi(x,t=0) \equiv \Phi_0 (x) $}  in the vicinity of the jump points,  
 while  the remainder  $ \Psi (x,t)$ 
 is expected to  
 be a regular function vanishing for $t=0$.
As a  result, we get   the following equation: 
  \begin{align}
& \! \!  \! \frac{\partial{\Psi  (x,t) }}{\partial t}    
  =    \int_0^1 \! \! V(x,y)  \left [\Psi(y,t)  -
   \Psi(x,t) \right ] dy +v(x) \Psi (x,t)   \nonumber 
 \\ & 
 +    \int_0^1 V(x,y) \,  \left [\Phi (y,t)  -
\Phi(x,t) \right ] \, dy
  -   w(x) \Phi (x,t) 
  \,  . 
  \label{psievol}
 \end{align}
 \begin{widetext}
 This is an inhomogeneous   evolution equation for  $\Psi (x,t)$,
 with starting  condition $\Psi (x,t=0)=0$.
For  its derivative at  $t=0$ we have 
   \begin{align}
\left.   \frac{\partial{\Psi  (x,t) }}{\partial t} \right   |_{t=0} =
&   \int_0^1 V(x,y) \,  \left [\Phi_0 (y)  -
\Phi_0(x) \right ] \, dy 
 -  w(x) \Phi_0 (x)  \equiv \chi (x) 
  \,  .
 \end{align}

  To avoid singularities   at the jump points,
 we should adjust $w(x)$ in such a way as to make 
$ \chi   (x)$ a continuous function of $x$. 
Then  $\Psi (x,t) \approx t \chi (x)  $ for small $t$.  The corrections to this approximation
can be found  by iterations. Namely, we 
represent $\Psi (x,t) = \sum_{n=1} \Psi_n (x,t)$ and start with 
  \begin{align}
& \Psi_1  (x,t)   
  \equiv     \int_0^t  d\tau \left \{    \int_0^1 V(x,y) \,  \left [\Phi (y,\tau )  -
\Phi(x,\tau) \right ] \, dy
  -   w(x) \Phi (x,\tau) \right \}  \  , 
  \label{psi1}
 \end{align}
generating further  terms using 
  \begin{align}
& \Psi_{n+1}  (x,t)   
  \equiv     \int_0^t  d\tau \left \{    \int_0^1 V(x,y) \,  \left [\Psi_n (y,\tau )  -
\Psi_n (x,\tau) \right ] \, dy
  +  v(x) \Psi _n (x,\tau) \right \}  \  .
  \label{psi11}
 \end{align}
 Since the derivative  of $\Psi_1 (x,t)$ for $t=0$ is given by $\chi (x)$,  we can  write
\begin{align}
& \Psi_1  (x,t)   
=   t \, \chi(x)+    \int_0^t  d\tau  \int_0^1 V(x,y) \, \left[\delta \Phi (y,\tau )  -
\delta \Phi(x,\tau)\right] \, dy
  -   w(x)\delta  \Phi (x,\tau)   \equiv  t \, \chi(x)+  \delta \Psi_1 (x,t)  \  . 
  \label{deltapsi1}
\end{align}
Here, $\delta  \Phi (x,\tau)   \equiv  \Phi (x,\tau)  - \Phi_0 (x)$ is the deviation of the Ansatz function  $\Phi (x,\tau)$
from its $\tau=0$ shape. 
For small $\tau$, the function $\delta  \Phi (x,\tau)$ has a rather sharp behavior 
at the jump points $x= \zeta_i$,   and this results in a rather sharp
behavior of $\delta \Psi_1 (x,t)$ at these points. Since each  iteration $\Psi_{n+1}  (x,t) $
is  generated linearly from a  previous $\Psi_{n} (x,t)$ one 
 (see Eq. (\ref{psi11}) ),   it  makes sense to
 split $\Psi (x,t)$  into a  ``smooth''  part $\Psi_\chi  (x,t)$ generated by  iterations  
 of $t \chi (x)$ and the remainder $\delta \Psi (x,t)$ generated by  iterations 
 of  $\delta \Psi_1 (x,t)$.   Thus, we have 
 \begin{align}
  \varphi(x,t) = \Phi(x,t) +  \delta \Psi (x,t) + \Psi_\chi (x,t) \ ,
 \label{ansatz2a}
 \end{align}
 where the first two terms, $\Phi (x,t)$ and $ \delta \Psi (x,t)$    have a rather sharp behavior 
 at the jump points  for small $t$, while   $\Psi_\chi (x,t) $ has a smooth behavior.

\section{Structure of Two-Photon Generalized Distribution Amplitude}

 In the lowest QCD order, the  non-singlet  two-photon GDA is given by \cite{Beiyad:2008ss} 
 \begin{align}
 \label{H0}
 \psi^q(x,\zeta,Q^2) &= \frac{N_C\,e_{q}^2}{2\pi^2} \log{\frac{Q^2}{m^2}}\, 
 \varphi (x,\zeta) \  ,
 \end{align}
 where the function $\varphi (x,\zeta) $  is proportional
to the $VV \to VV$ component 
\begin{align}
\label{H1}
\varphi (x,\zeta) &=
\frac{\bar{x}(2x-\zeta)}{\bar{\zeta}}\theta(x-\zeta)+\frac{\bar{x}(2x-\bar{\zeta})}{\zeta}\theta(x-\bar{\zeta}) 
-\frac{x(2\bar x-\bar \zeta)}{\zeta}\theta(\zeta-x)- \frac{x(2\bar x- {\zeta})}{\bar{\zeta}}\theta(\bar{\zeta}-x)
\end{align}
of the ERBL evolution kernel matrix. QCD corrections  induce further evolution of the photon GDA.
Namely, its derivative with respect to $\ln Q^2$   obeys 
 ERBL evolution equation with the  $qq\to qq$ kernel considered above.
In what follows, we study  ERBL  evolution of the function 
 $ \varphi (x,\zeta,t)$ which for the starting evolution point 
 $t=0$ coincides with $ \varphi (x,\zeta)$.

\begin{figure}[htb]
\includegraphics[width=7.8cm]{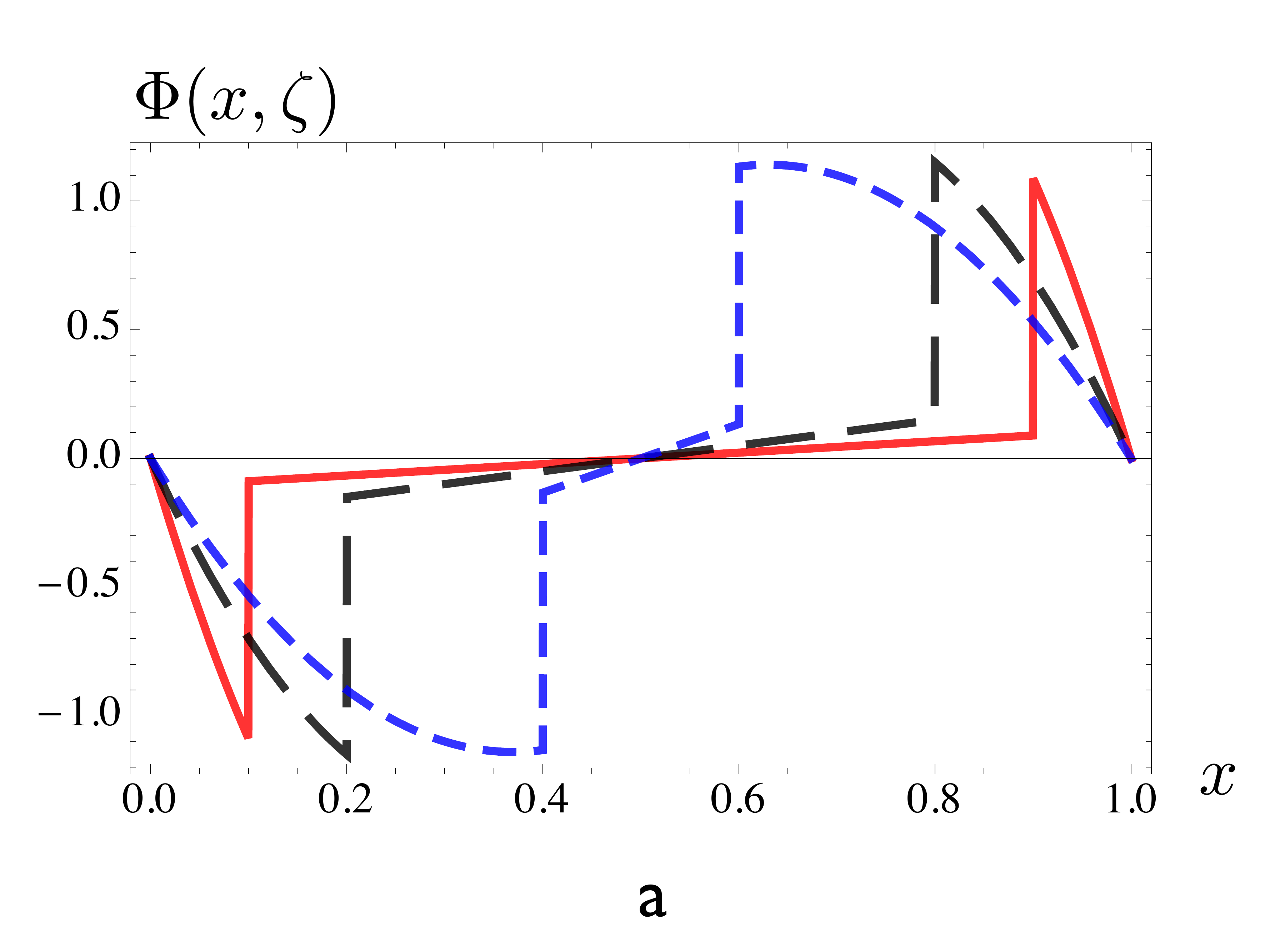}
\vspace{-2mm}
\includegraphics[width=7.7cm]{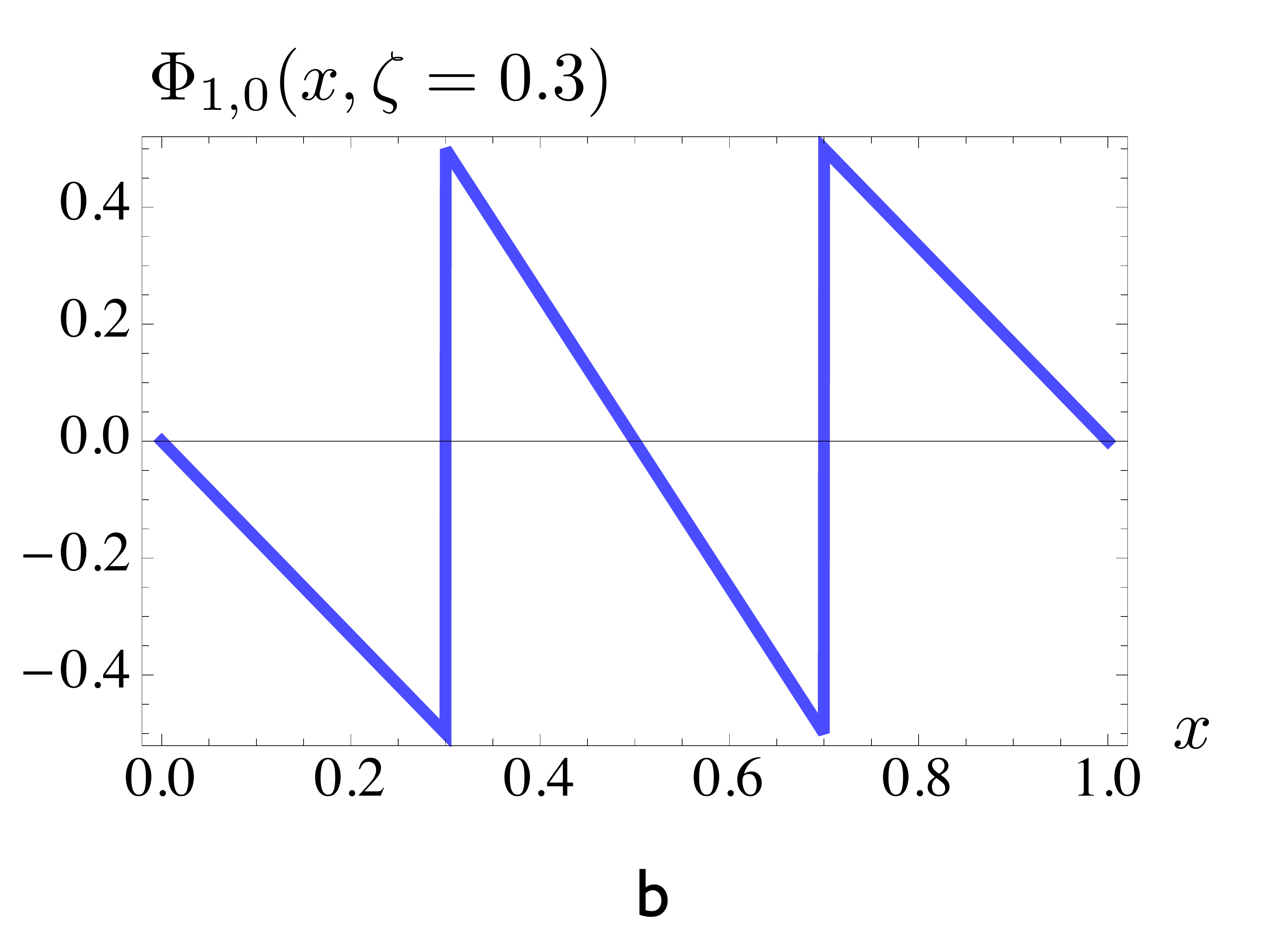}
\caption{ a)Two-photon  GDA profile function $\Phi(x,\zeta)$ at
values  $\zeta=0.1;  0.2; 0.4$.  b) The $x$-profile of the jump part 
$\Phi_{1,0}(x,\zeta)$ at $\zeta=0.3$.  }
\label{twoph}
\end{figure}


The  function $ \varphi (x,\zeta)$ is  antisymmetric with respect to $x \leftrightarrow \bar x $ interchange 
and   symmetric with respect to $\zeta  \leftrightarrow \bar \zeta $ interchange.
Thus, without loss of generality, we may choose $\zeta \leq 1/2$.
Then $0 \leq \zeta \leq \bar \zeta \leq 1$, and it makes sense to  
explicitly write the function in 
each of the three regions:
\begin{align}
\label{H11}
\varphi (x,\zeta) =& \left\{  \frac{x}{\zeta \bar \zeta}  \left [ \bar \zeta^2 + \zeta^2 - 2  \bar x \right ] 
\theta(0\leq x\leq \zeta) 
-  \frac{\zeta (1-2 x)  }{1- \zeta} 
\theta(\zeta \leq x \leq \bar{\zeta}) 
- \frac{\bar x}{\zeta \bar \zeta}  \left [ \bar \zeta^2 +   \zeta^2 - 2 x  \right ] 
\theta(\bar \zeta \leq x\leq 1)   \right\}\,.
\end{align}
The  function is discontinuous at $x=\zeta$ and $x=\bar{\zeta}$. Fig.\ref{twoph}a
  shows the $x$-profile of the two-photon  GDA at different $\zeta$ values.
As $x$ approaches $\zeta$, the  limiting value  of the function from the left
is 
\mbox{$\varphi (\zeta_-,\zeta)= -1-\zeta \, \frac{1-2 \zeta}{1-\zeta} $}, 
while from the right we have
\mbox{$\varphi (\zeta_+,\zeta)= -\zeta \, \frac{1-2 \zeta}{1-\zeta} $}, 
so that the jump
$\varphi (\zeta_+,\zeta)- \varphi (\zeta_-,\zeta) = 1$
is equal to 1.
According to our discussion in the preceding section,
to treat the evolution of a  jump, we should represent the initial function 
  as a sum of a 
function $ \varphi_{2}(x,\zeta)$  continuous   in the vicinity  of each jump, and a function
 $\varphi_{1}(x,\zeta)$ 
that has an antisymmetric  jump of 
 necessary size. 
 The function $\varphi_{1}(x,\zeta)$ will also specify the initial 
 form of the $\Phi$-part of the evolution Ansatz (\ref{ansatz2})
 for this function, so we  will denote
 it as $\Phi_{1,0} (x,\zeta)$. For simplicity, we will choose   it 
 to be given by  linear functions of $x$ in 
each of the three regions. As a result,
\begin{align}
\Phi_{1,0}(x,\zeta)=&-\frac{x}{2\zeta}\theta(0<x<\zeta)+
\frac{\bar{x}}{2\zeta}\theta(\bar{\zeta}<x<1)  
 + \frac{1-2x}{2(1-2\zeta)}\theta(\zeta <x <\bar \zeta) \  . 
\end{align}
The function $\Phi_{1,0}(x,\zeta)$  is discontinuous at $x=\zeta$ and 
\mbox{$x=\bar{\zeta}$}, see Fig.\ref{twoph}b,   
 where it is shown for 
 $\zeta =0.3$.
The function $\Phi_{2,0}  (x,\zeta)$ specifying the initial shape
of the continuous part  is obtained as the difference 
between $\varphi (x,\zeta)$ and $\Phi_{1,0} (x,\zeta)$.

  \section{Evolution of the jump part of  two-photon GDA}

Iteration of the initial function $\Phi_{1,0}  (x,\zeta)$ with evolution kernel  gives
\begin{align}
  \Phi_{1,1} (x,\zeta)\equiv &  \int_0^1 V(x,y)[\Phi_{1,0} (y,\zeta)-\Phi_{1,0} (x,\zeta)] \, dy 
 =
{\theta(0<x<\zeta<1/2)}
\biggl\{ (1-4\zeta)\, \frac{x}{2 \zeta} 
 \nonumber\\ 
   &
   + \frac1{2 \zeta}\, [ x \bar x \ln x +(1-x\bar x)\ln\bar{x}]
    - \ln(\zeta-x)- \ln(\bar{\zeta}-x) +
  \frac{  \ln \zeta 
 +  \ln\bar{\zeta} \,  (1-3\zeta)/ \zeta }{2 (1-2\zeta) }\,  \bar x
  \nonumber\\
 &
   -  \frac{1-4 \zeta}{ \zeta (1-2\zeta) } \left [(x-\zeta )  \ln(\zeta-x)+ 
  (\bar \zeta -x) \ln(\bar{\zeta}-x)
  \right ]
 \biggr\} -\{x\rightarrow \bar{x}\}
 \nonumber\\
 &+{\theta(\zeta<x<\bar \zeta )} \biggl\{ \frac{1- 2 x}{2  (1-2\zeta)}
  \Bigl  [(1-4\zeta)
+ \ln\bar{\zeta} \, (1-3\zeta)/\zeta \Bigr ]
  + \ln(x-\zeta)- \ln(\bar{\zeta}-x) 
\nonumber \\ &
+ \frac{1-4 \zeta}{2 \zeta (1-2\zeta)}
\Bigl  [(x- \zeta )\ln(x-\zeta  ) - (\bar{\zeta}-x) \ln(\bar{\zeta}-x)
+\bar x \ln\bar{x} -x \ln x \Bigr ]  +
\frac{x^2 +\bar x^2}{2  (1-2\zeta)}
 \ln \left ( \frac{\bar{x}}{ x} \right ) 
\biggr\} \  . 
\label{Phi11}
\end{align}

\end{widetext}

As expected,  $\Phi_{1,1} (x,\zeta)$ has logarithmic singularities
 \begin{align}
- \ln(\zeta-x)\theta(x<\zeta)
+  \ln(x-\zeta)&
\theta \left (\zeta<x<1/2\right )
 \nonumber\\ &
-  \{ x \to \bar x \} 
\end{align} 
 for  $x=\zeta$ and $x=\bar{\zeta}$ (see Fig.\ref{phi11}a).
The  sum of these terms  may be written as 
 $2(\ln |x-\zeta|+ \ln |x-\bar \zeta|)\Phi_{1}(x,\zeta)$ 
plus regular terms, which suggests to take the Ansatz (\ref{ansatz2})
with $w(x)$ containing $2(\ln |x-\zeta|+ \ln |x-\bar \zeta|)$.
Namely, let us try  the function $ w_0(x,\zeta)$   given by
\begin{align} 
 w_0(x,\zeta) =   4+2\ln |x-\zeta|+2\ln |x-\bar \zeta| \ .
\end{align}
The constant part ``4''  was chosen  to make the integral
of \mbox{$ w_0(x) $} closer to zero  
(it vanishes both for $\zeta=0$ and $\zeta=1$), i.e. to keep  the overall normalization 
of the Ansatz factor closer to 1.  
Resulting  function 
(which gives the first term of the $\Psi$-part of the Ansatz (\ref{ansatz2})) is given by 
\begin{align}
\Psi_{1,1}^{(0)} (x,\zeta)=\Phi_{1,1} (x,\zeta) 
- w_0 (x, \zeta) \Phi_1 (x, \zeta) 
\label{Phi11mod}
\end{align}
 and  shown  in Fig.\ref{phi11}b.
 One can see that, after the subtraction of singularities, we still have finite jumps
for $x=\zeta$ and $x= \bar \zeta$.  
Explicit calculation gives
\begin{align}
\Psi_{1,1}^{(0)}(\zeta_+,\zeta) & - \Psi^{(0)}_{1,1} (\zeta_-,\zeta)  = 
4+2\ln(|1-2\zeta|)  \nonumber \\
 & 
+(2-\zeta) \ln\zeta+(2- \bar \zeta)\ln \bar{\zeta}
 \equiv 
- w_1 (\zeta) \ .
\end{align}
Adding $w_1 (\zeta) \Phi_{1,0}(x,\zeta) $ to $\Psi_{1,1}^{(0)}(x,\zeta)$,
we obtain the correction function  $\Psi_{1,1}(x,\zeta) \equiv \chi (x,\zeta)$
that is continuous at the border points
$x=\zeta$ and $x=\bar \zeta$ (see Fig.\ref{phi11}c). 

\begin{widetext}

 \begin{figure}[htb]
 \centering
\includegraphics[width=5.5cm]{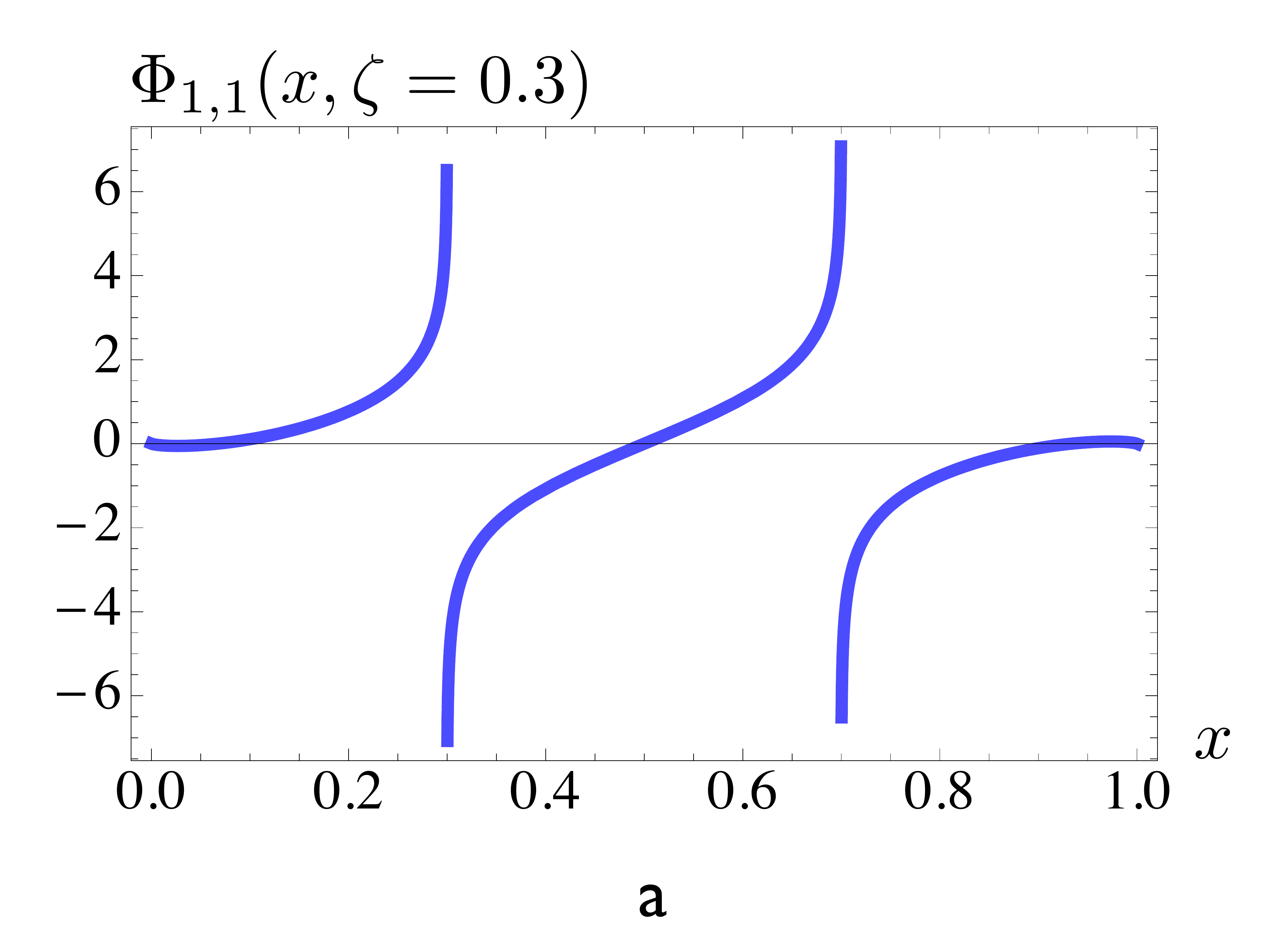}
\includegraphics[width=5.5cm]{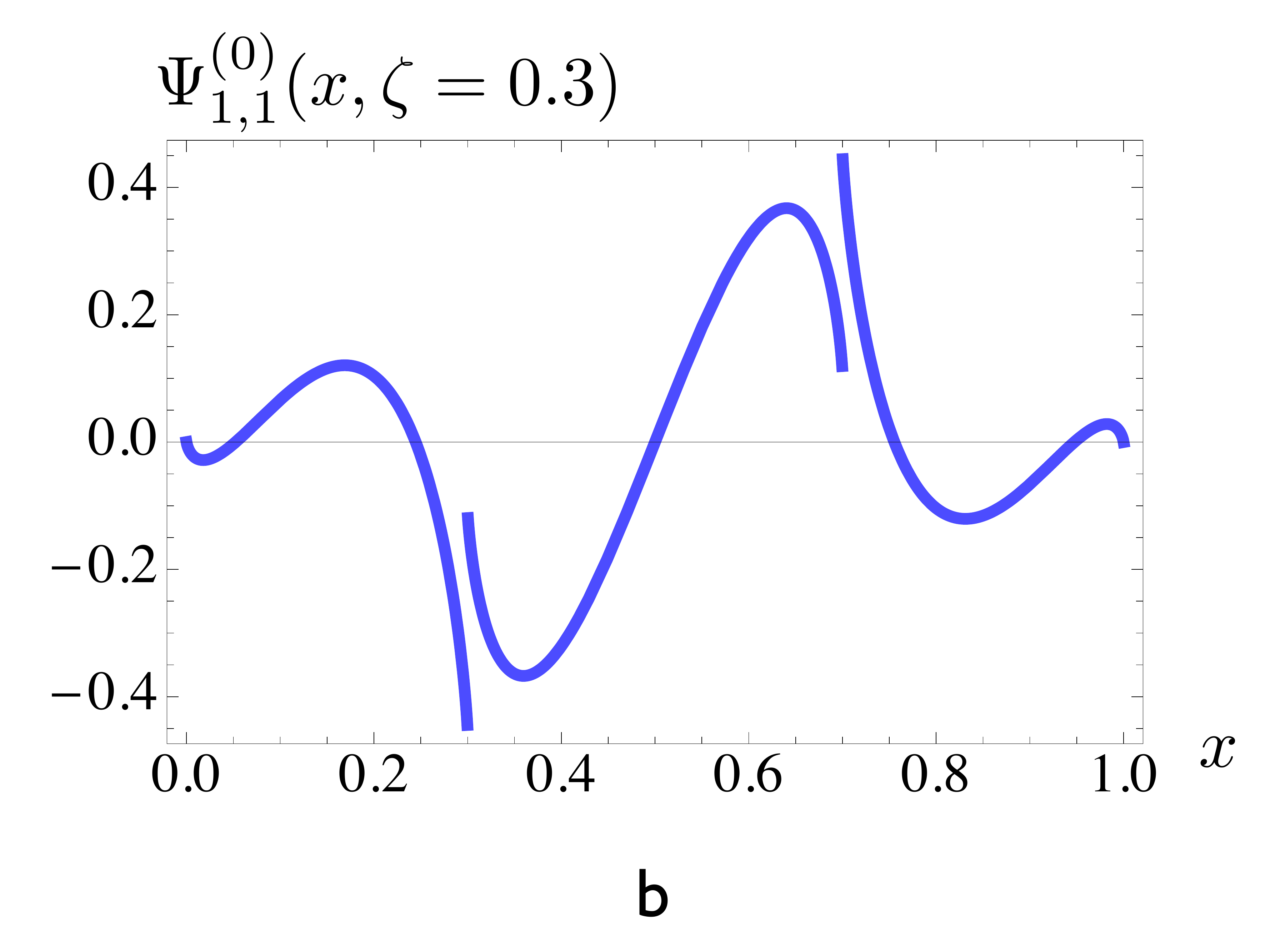}
\includegraphics[width=5.5cm]{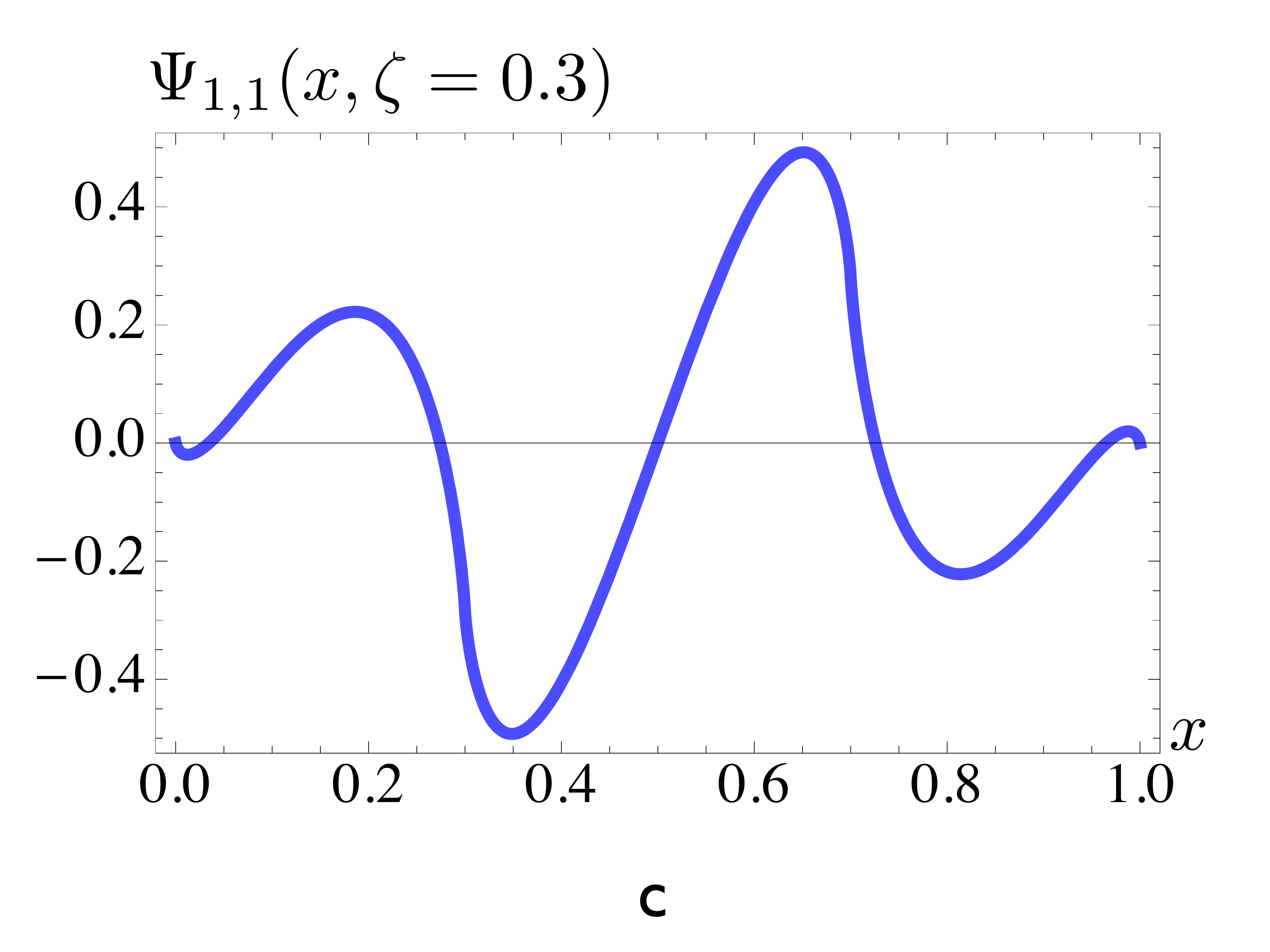}
\vspace{-10pt}
\caption{a)  First iteration $\Phi_{1,1} (x,\zeta)$  for 
$\zeta =0.3$.  b)  Correction functions $\Psi_{1,1}^{(0)}(x,\zeta=0.3)$  with $w_0$ Ansatz.  c)  Correction functions $\Psi_{1,1}^{(0)}(x,\zeta=0.3)$ 
with modified Ansatz.}
\label{phi11}
\end{figure}

This corresponds to  the following  $\Phi$-part of the  Ansatz (\ref{ansatz2}) 
\begin{align}
\Phi_{1} (x,\zeta,t)=& e^{t v(x)} \left (\frac{|1-x/\zeta||1-x/\bar \zeta|}{|1-2\zeta|}
\right )^{2t}  
\zeta^{t\zeta} \bar \zeta^{t \bar \zeta} 
\Phi_{1,0}  (x, \zeta) 
\end{align}
for  the function $\varphi_1 (x,\zeta,t)$.  The function  $\Phi_1(x,\zeta,t)$  is
illustrated  in Fig. \ref{phiansatzmod02}a.

\begin{figure}[htb]
\centering
\includegraphics[width=6cm]{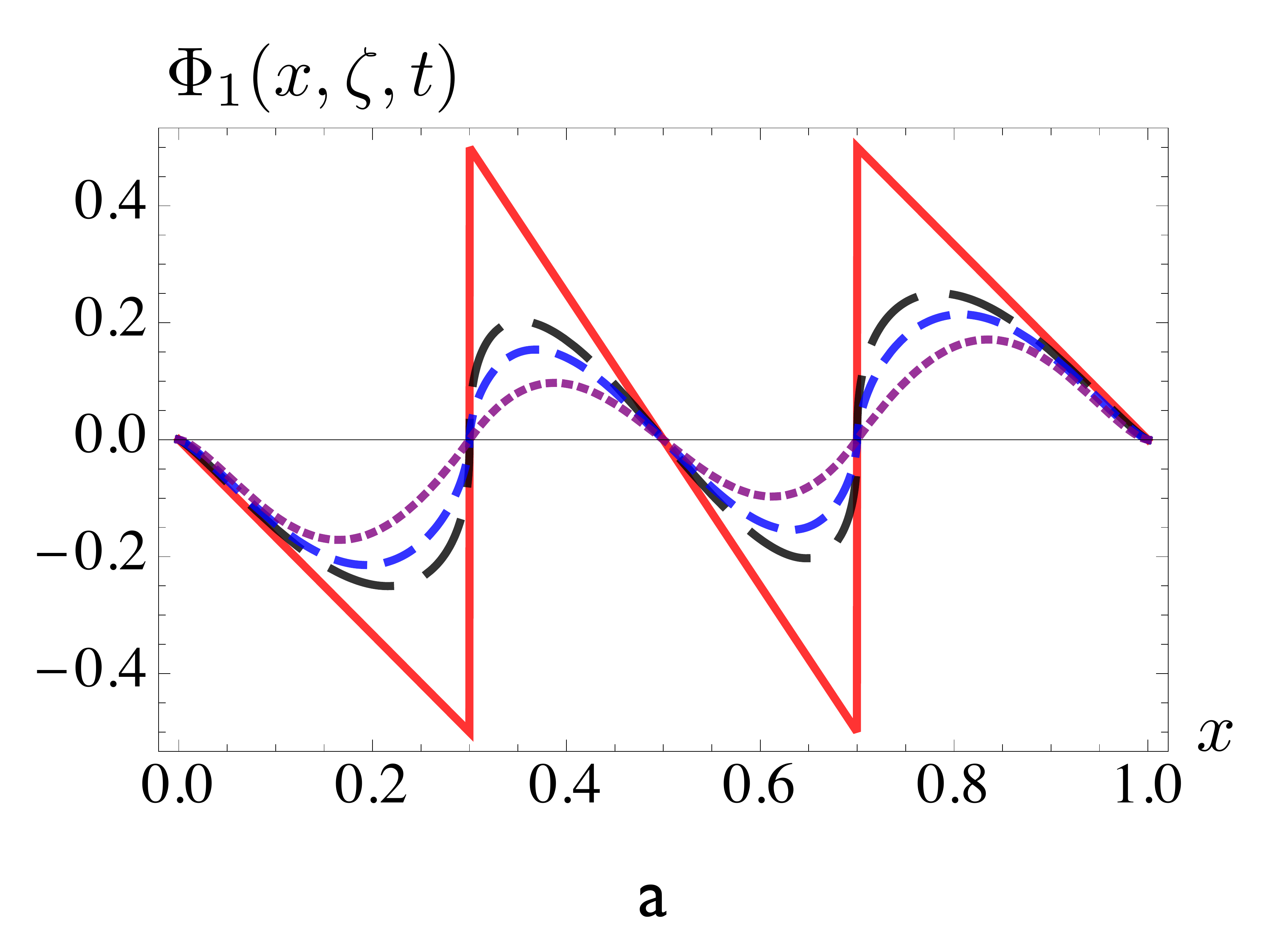}
\includegraphics[width=5.7cm]{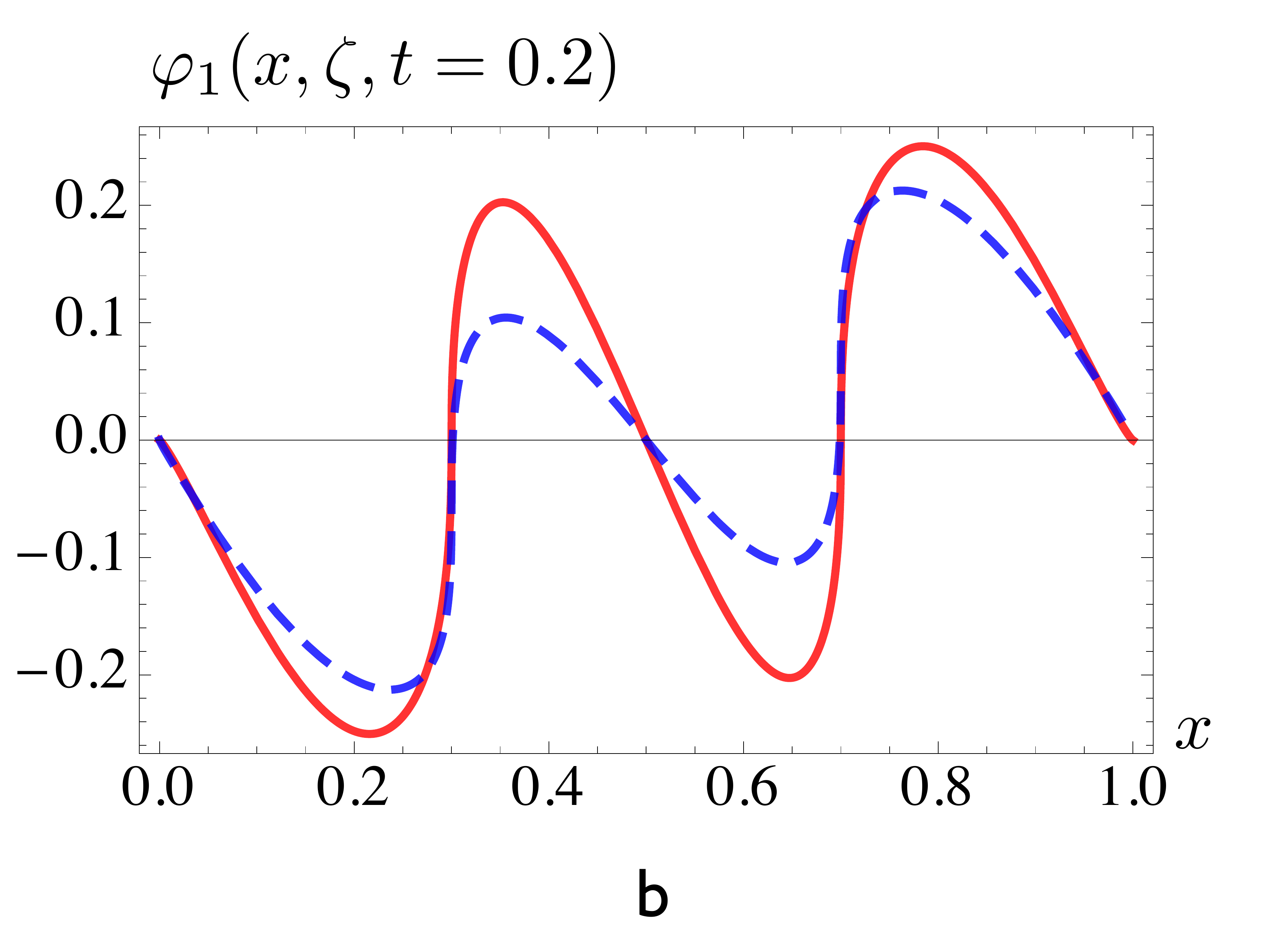}
\includegraphics[width=5.7cm]{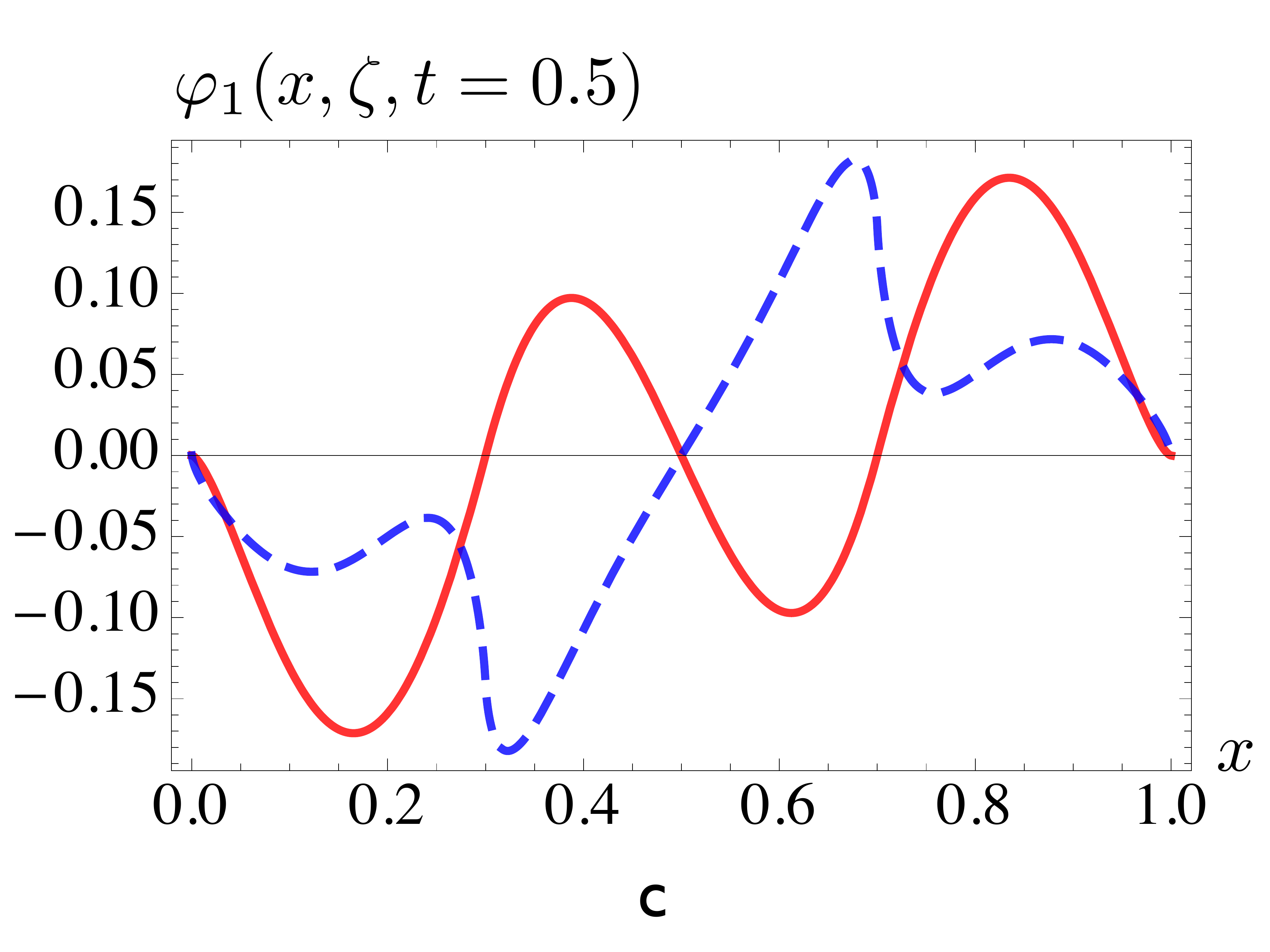}
\caption{a) Ansatz function  $\Phi_1 (x,\zeta=0.3, t )$  for $t=0, \ 0.2, \ 0.3, \ 0.5$.
Effect of inclusion of $\Psi_{1,1}(x,\zeta)$ correction for $t=0.2$ (b) and
$t=0.5$ (c); $\zeta =0.3$ in both cases.  }
\label{phiansatzmod02}
\end{figure}

According to Ansatz (\ref{ansatz2}), after fixing the function $w(x)$ 
from the requirement of continuity of $\chi (x,\zeta)$,
one should deal  with the evolution equation (\ref{psievol}) 
for the $\Psi$-part of the  Ansatz. 
This equation specifies that $\partial \Psi_1 (x,\zeta,t)/\partial t$ 
for $t=0$ is given by  $\chi(x,\zeta)$. 
Thus, for small $t$, we can approximate $\Psi_1 (x,\zeta,t)$ 
by $t \chi (x,\zeta)$. As one can see from 
Fig.\ref{phiansatzmod02}b,  the correction due to the $\Psi$ term 
 is rather small 
for $t=0.2$. It just reduces somewhat the amplitude of oscillations.

However, 
the correction  becomes more and more  visible with growing $t$, 
see Fig.\ref{phiansatzmod02}c, where the evolved function is shown
for $t=0.4$  with 
and without the first  $\Psi$-type correction term included.
The total function is now clearly nonzero 
at the ``border''  points $x=\zeta$ and $x = 1-\zeta$.
This is because $ \chi (x,\zeta)$ is nonzero at these points. 
As we discussed, the $\Psi$-part becomes dominant for large $t$
and brings the shape  of $\varphi_1 (x, \zeta,t)$ to the 
 asymptotic form 
$x\bar x (2x-1)$ of the antisymmetric DAs.
We can see that, for $t=0.4$ already,   the total 
function  resembles the asymptotic shape 
$x\bar x (2x-1)$. 
However, for such large $t$ values the simplest 
linear-$t$ approximation for  $\Psi (x,\zeta,t)$ is too crude,
and one should go beyond the first iteration.

As argued in the discussion after Eq. (\ref{ansatz2}), it makes sense to
split $\Psi (x,,\zeta,t)$ into a part generated by iterations of $t \chi (x, \zeta)$,
and the remainder $\delta \Psi (x,,\zeta,t)$ given by iterations 
of the terms reflecting the deviation $\delta \Phi (x,\zeta;t) \equiv 
\Phi (x,\zeta;t) -\Phi_0 (x,\zeta)$ of the Ansatz function $\Phi (x,\zeta;t)$
from its $t=0$ form $\Phi_0 (x,\zeta)$.  The starting term
$\delta \Psi_1 (x,,\zeta,t)$ has a sharp behavior at the jump points 
of $\Phi_0 (x,\zeta)$ (see Fig. \ref{phideltapsi}a), acquiring an  infinite slope there as  $t\to 0$. 
Up to this point of the calculation, all iterations are calculated analytically, 
however the next iteration, $\delta \Psi(x\zeta,t)$   (see Fig. \ref{phideltapsi}b) was  calculated numerically.

\begin{figure}[htb]
\centering
\includegraphics[width=5.7cm]{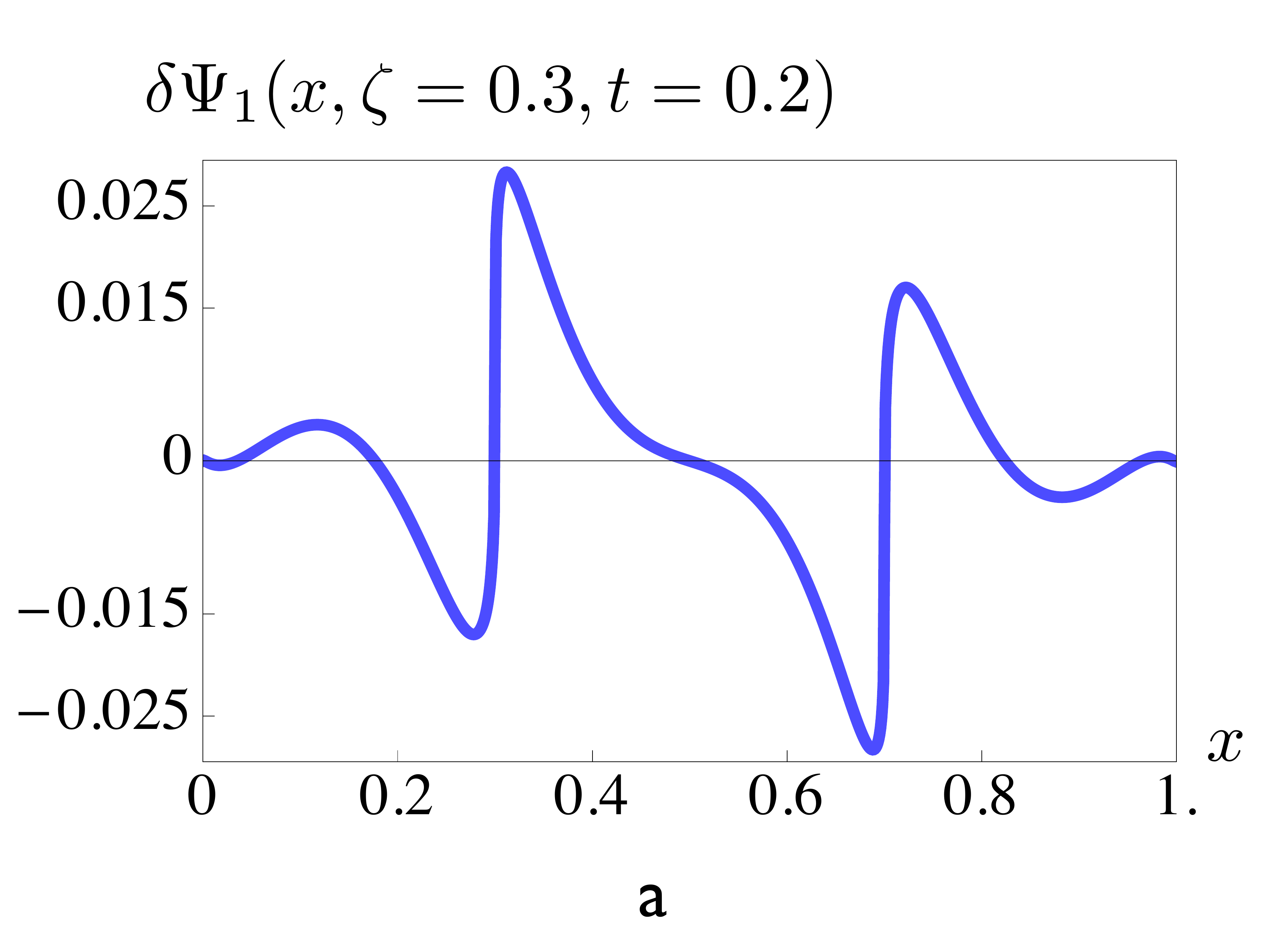}
\includegraphics[width=5.7cm]{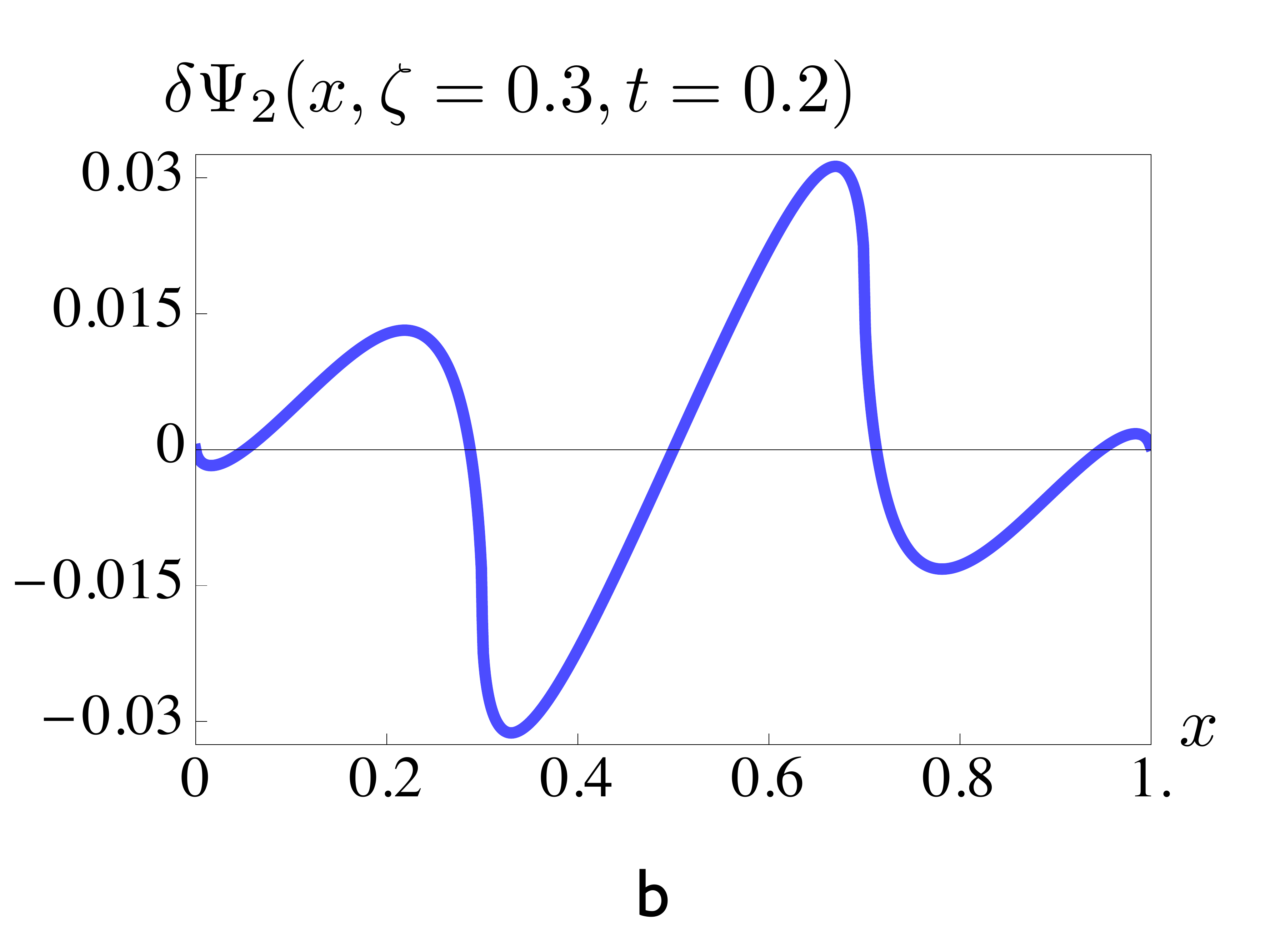}
\caption{a) $\delta \Psi_1(x,\zeta,t)$. b) $\delta \Psi_1(x,\zeta,t)$  ($\zeta=0.2$ and $t=0.2$).  }
\label{phideltapsi}
\end{figure}

 The amplitudes of both $\delta\Psi_1$ and $\delta\Psi_2$ is very small compared to the amplitude of $\Phi(x,\zeta,t)$. As suggested in the previous section $\Psi(x,\zeta,t)\approx t \chi(x,\zeta)$ is a good approximation. In Fig. \ref{phideltapsitoget}, the function $\Phi(x,\zeta,t)$ with $\delta\Psi_1(x,\zeta,t)$ (a) 
 and $\delta\Psi_2(x,\zeta)$  9b) is depicted. One can see that the contribution from $\delta \Psi(x,\zeta,t)$ is negligible.

\begin{figure}[h]
\centering
\includegraphics[width=6cm]{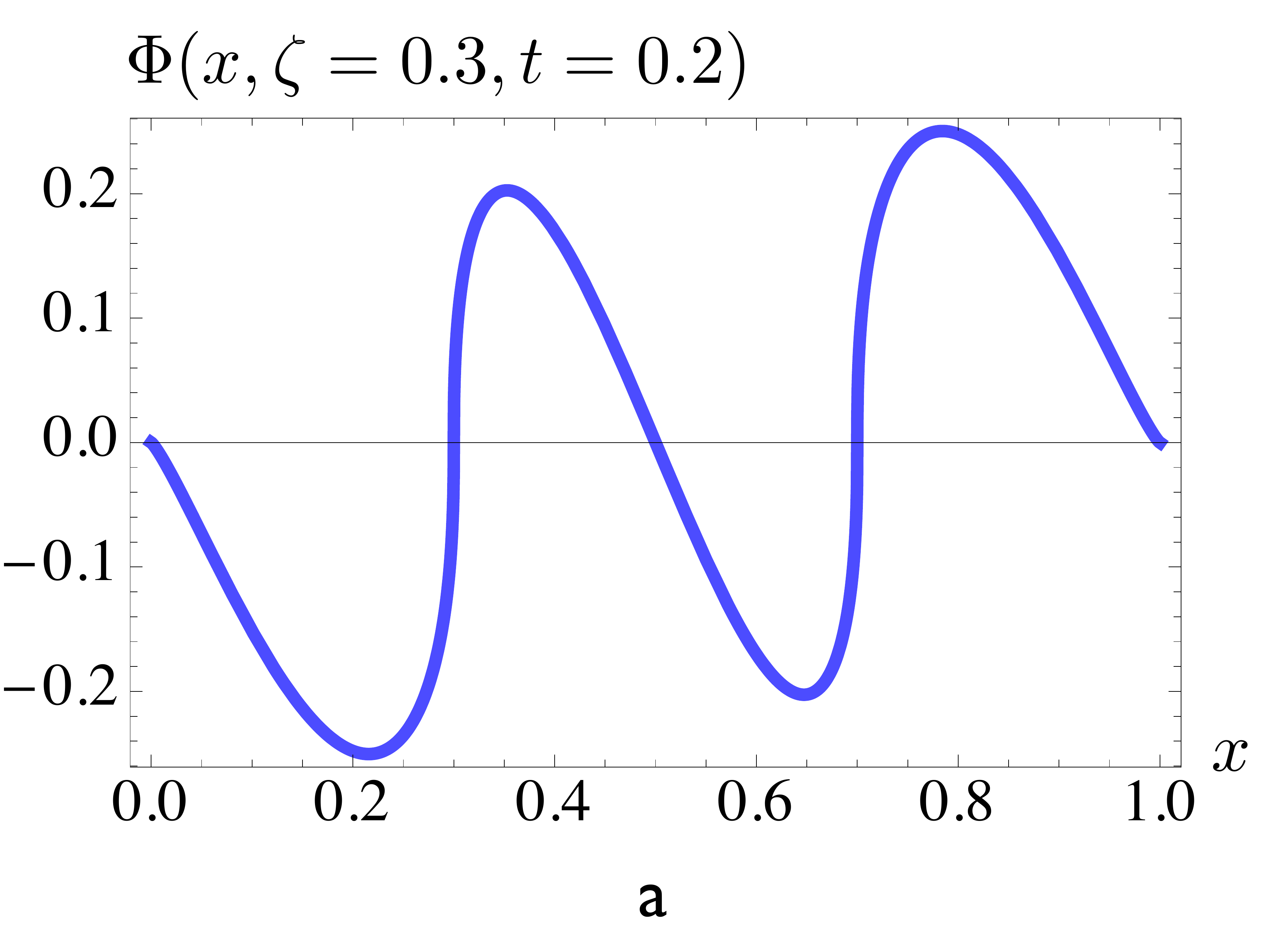}
\includegraphics[width=6cm]{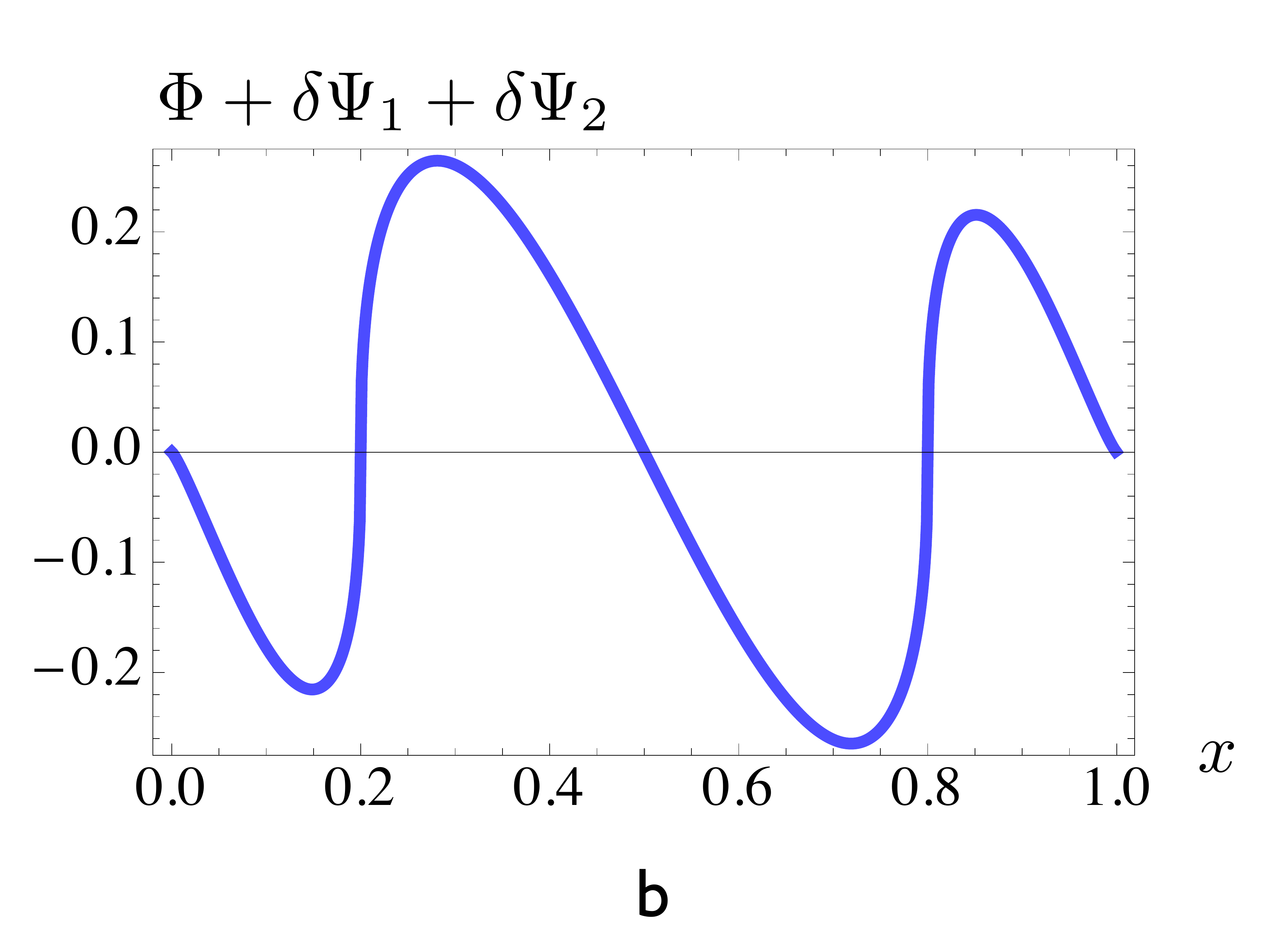}
\caption{a) $\Phi(x,\zeta,t)$. b) $\Phi(x,\zeta,t)$ and $\delta \Psi_1(x,\zeta,t)+\delta\Psi_2(x,\zeta,t)$
($\zeta=0.2$ and $t=0.2$).  }
\label{phideltapsitoget}
\end{figure}

\section{Evolution of  cusp part of two-photon GDA }

\subsection{Decomposition}

In this section, we study  evolution of the second function, namely $\varphi_{2}(x,\zeta)$. 
Its initial form  $\Phi_{2,0}(x,\zeta)$   is continuous for  $x=\zeta$ and $x=\bar{\zeta}$ and  
 is given by 
 \begin{align}
 \Phi_{2,0}(x,\zeta)=&-\left ( \frac{x}{\zeta} \right ) \, 
  \frac{1+\zeta -4 \zeta^2+4(\zeta-x) }{2(1-\zeta)}\, \theta(x< \zeta)
  - \{ x \to 1-x \} 
   -\left ( \frac{1-2x}{1-2\zeta} \right ) 
\frac{1+\zeta-4\zeta^2}{2(1-{\zeta})}\,
 \theta(\zeta< x < \bar{\zeta})
 \ ,   
  \end{align}
see   Fig.\ref{phi2_0}a.   
We can separate this function 
 \begin{align}
 \Phi_{2,0}(x,\zeta)=&-  
  \frac{1+\zeta -4 \zeta^2}{2(1-\zeta)}  \Phi^\text{L}_{2,0}(x,\zeta) +
   \Phi^\text{C}_{2,0}(x,\zeta) 
    \end{align}  
  into a term proportional to  a  linearized function,   Fig.\ref{phi2_0}b, 
  \begin{align}
 \Phi^\text{L}_{2,0}(x,\zeta)=& \frac{x}{\zeta} \, \theta(x< \zeta)- \{ x \to 1-x \} 
 -\left ( \frac{1-2x}{1-2\zeta} \right )
 \theta(\zeta<x<\bar{\zeta})
 \ , 
  \end{align}
and the remaining curvy part   (see Fig.\ref{phi21l}a below), 
 \begin{align}
  \Phi^\text{C}_{2,0}(x,\zeta)=& \frac{2x(x-\zeta)}{\zeta\bar{\zeta}}\, \theta(x<\zeta)- \{ x \to 1-x \} 
  \ .
   \end{align}

\begin{figure}[t]
\centering
  \includegraphics[width=5.5cm]{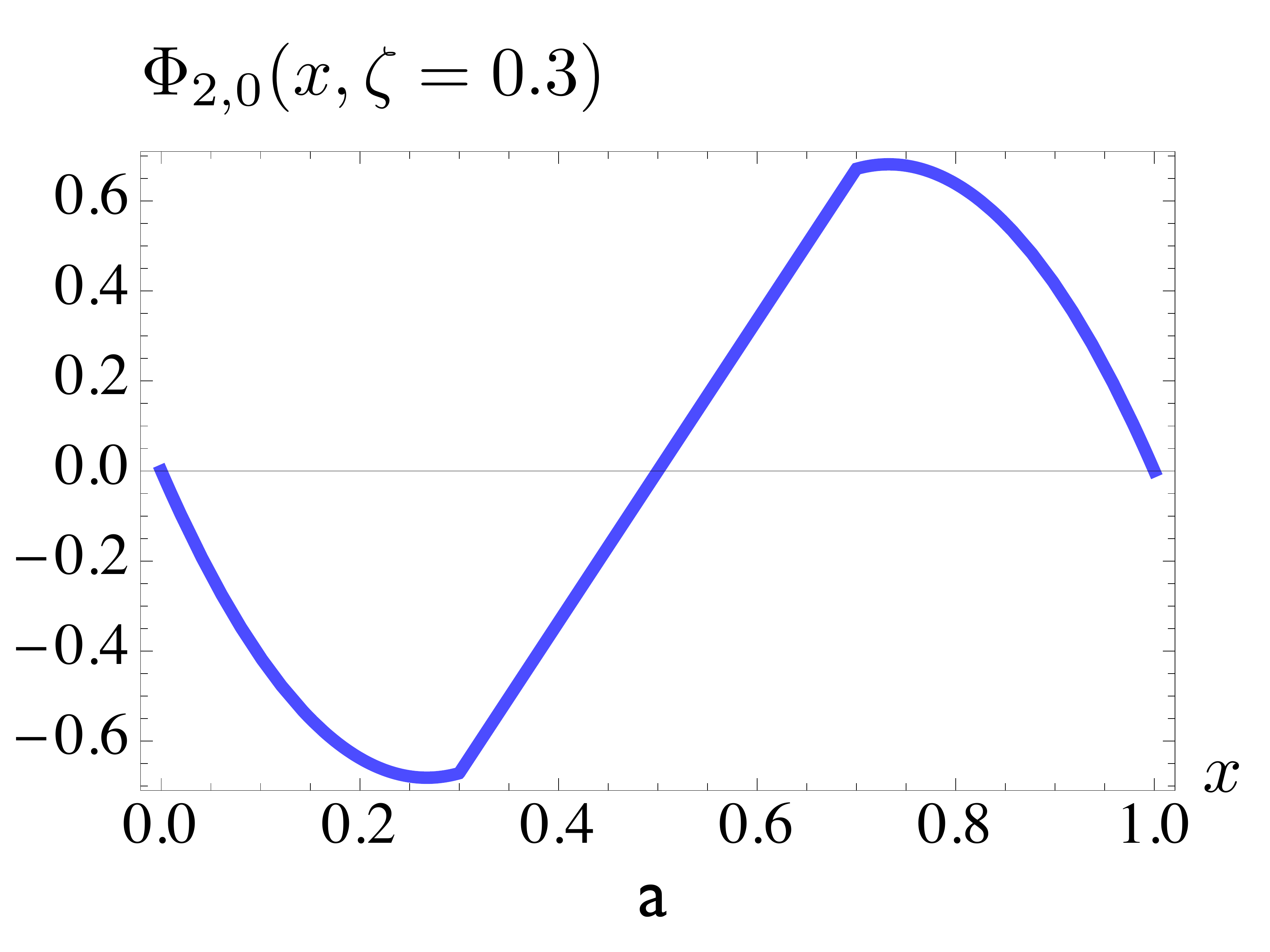}
    \includegraphics[width=5.5cm]{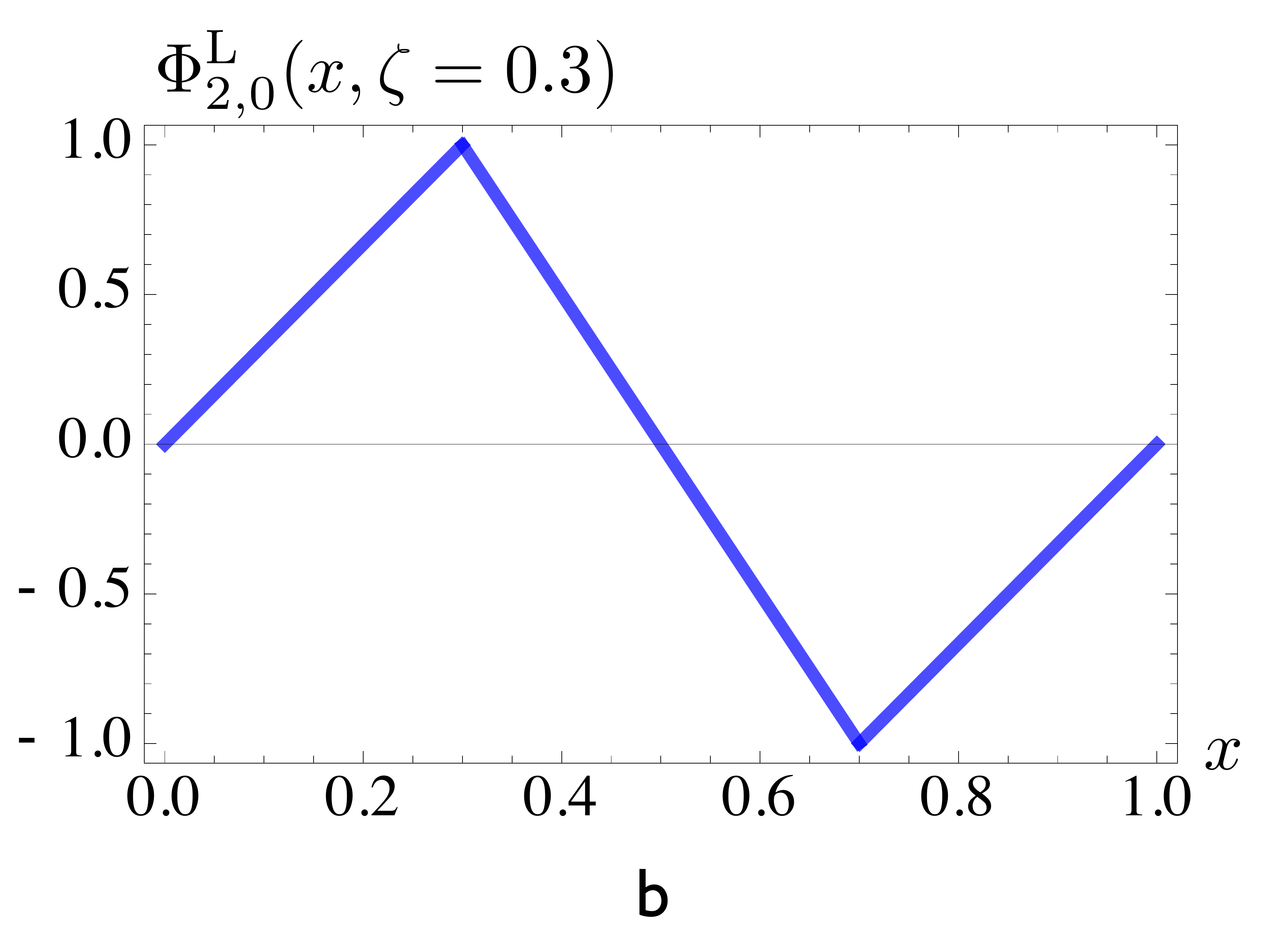}
        \includegraphics[width=5.5cm]{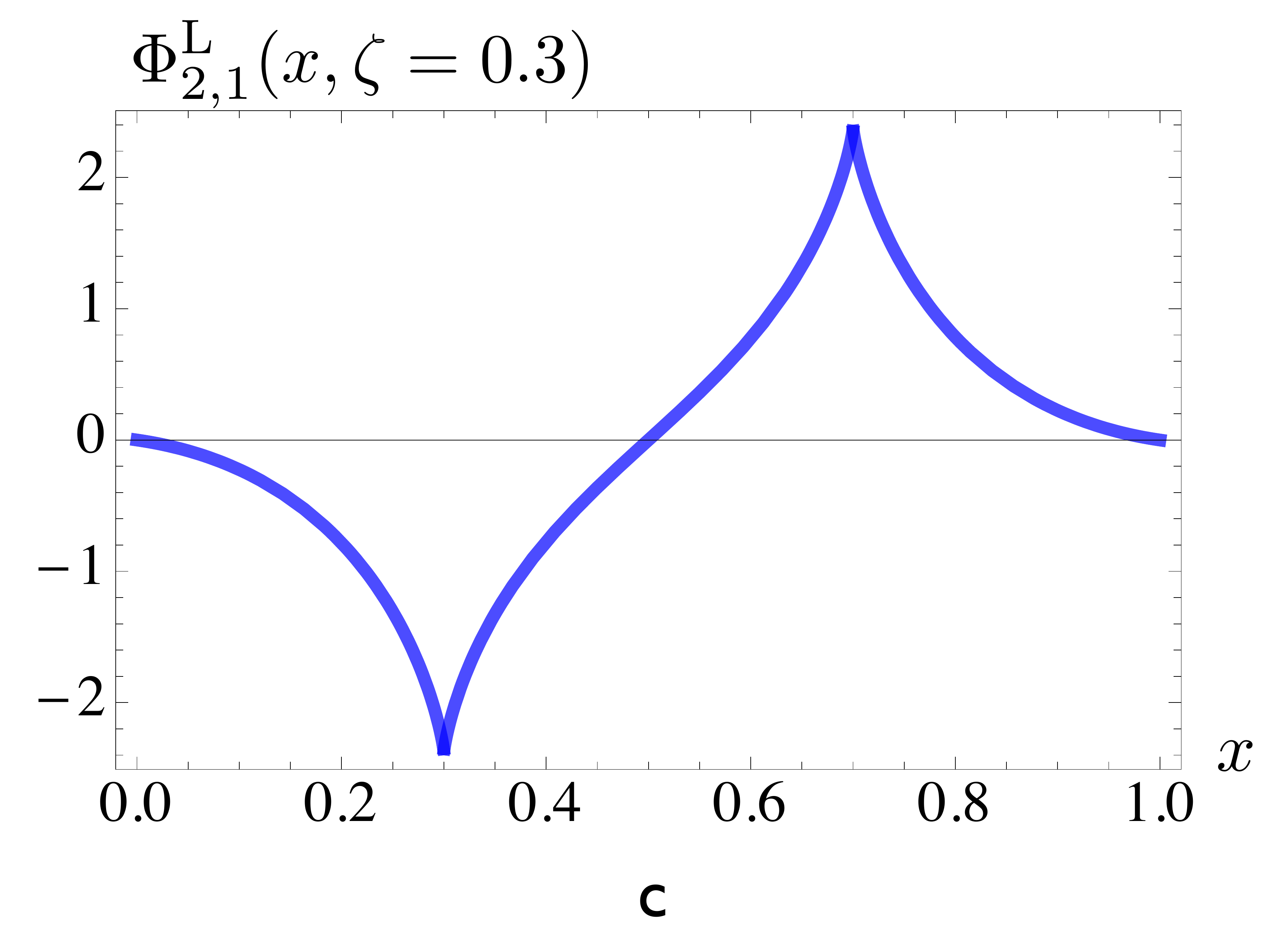}
\caption{a) Initial function  $\Phi_{2,0}(x,\zeta)$ for $\zeta=0.3$. 
b) Initial form $\Phi^L_{2,0}(x,\zeta=0.3)$  of the  linearized part.   
c)  First iteration function   $\Phi^L_{2,1}(x,\zeta=0.3)$    for the linearized part.
}
\label{phi2_0}
\end{figure}

\subsection{Evolution of linearized part}

Since   $\Phi_{2,0} (x,\zeta)$  is a continuous function vanishing at the end points,
the easiest way to get its  
evolution  is to use straightforward $t^n$ expansion
with coefficients given by  successive  iterations of the evolution 
kernel with $\Phi_{2,0} (x,\zeta)$. 
The first iteration of   $ \Phi^\text{L}_{2,0}(x,\zeta)$ gives   
 \begin{align}
     \Phi^\text{L}_{2,1}(x,\zeta)=
     &
     -\frac{\theta(0<x<\zeta)}{\zeta(1-2\zeta)}
     \biggl\{
     +{\bar x} \Bigl [ \bar \zeta  \ln \bar \zeta -   \zeta  \ln \zeta \Bigr ] 
      + (1-2\zeta)  \Bigl [x+ x\bar{x} \ln x +(1-x\bar x)\ln \bar x
\Bigr ]
     \nonumber\\ &
     + (\zeta-x)\ln(\zeta-x)-(\bar \zeta-x)\ln(\bar \zeta-x)     \biggr\}
     -\{x\rightarrow \bar{x}\}    
     \nonumber\\ &
   -   \frac{ \theta(\zeta<x< \bar \zeta)}{1-2\zeta }
    \biggl\{  (1-2x ) \Bigl [ 1 +\frac{\bar \zeta}{\zeta} \,  \ln\bar{\zeta}\Bigr ] 
    +2 x\bar x \ln \frac{\bar x} {x}  
     + \left ( \frac{x}{\zeta} -1 \right ) \ln \left (1-\frac{\zeta}{x} \right )  
     -
     \left (\frac{ \bar x } {\zeta}-1 \right )\ln \left (1-\frac{\zeta }{ \bar x}  \right )
    \biggr\}  
     \nonumber\\ &
+ v(x) \Phi^\text{L}_{2,0} (x,\zeta) 
\, .
 \end{align}
Here, as usual,  $v(x)$ is $3/2+x\ln \bar{x}+\bar{x}\ln x$.   
   The structure of the result  is very similar to  that of  $\Phi_{2,1}(x, \zeta)$.
   However, the potentially singular logarithmic terms $\ln |x -\zeta|$
   and \mbox{$\ln |x -\bar \zeta|$}  are accompanied in this case by 
   $(x -\zeta)$ or  $(x -\bar \zeta)$  factors, respectively,
   and vanish at these points, though having singular derivatives  there.
  Thus, the function  $\Phi_{2,1}(x, \zeta)$, shown in  Fig. \ref{phi2_0}c, 
  is continuous     at points $x=\zeta$ and $x=\bar{\zeta}$.

\subsection{Evolution of curvy part and total result} 

Initially, the support   region for the curvy part  $\Phi^\text{C}_{2,0}(x,\zeta)$
is restricted by two segments $0<x<\zeta$ and $\bar \zeta <x <1$, see  Fig.\ref{phi21l}a.
   \begin{figure}[htb]
 \centering
           \includegraphics[width=6.25cm]{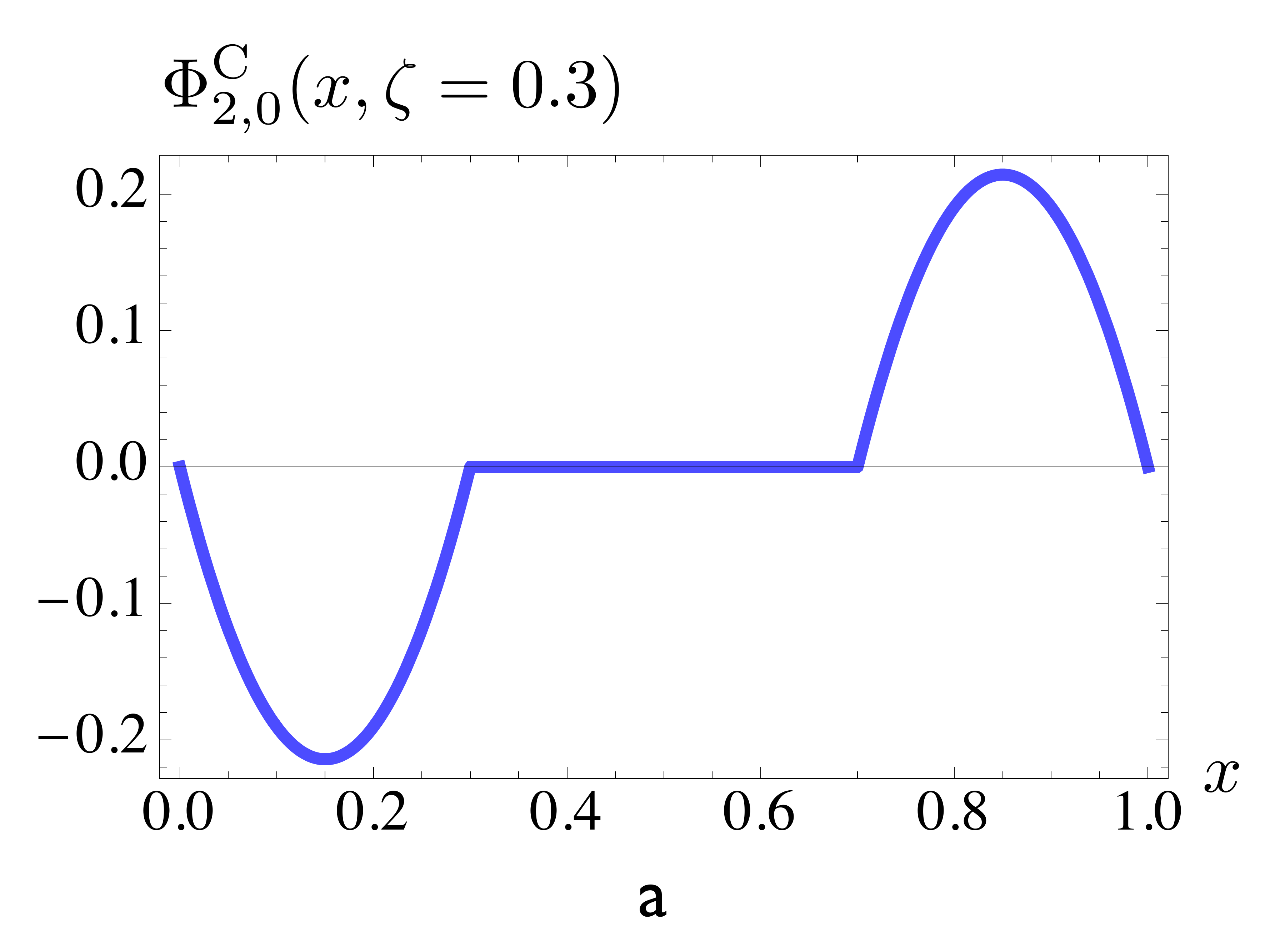} 
         \includegraphics[width=6.25cm]{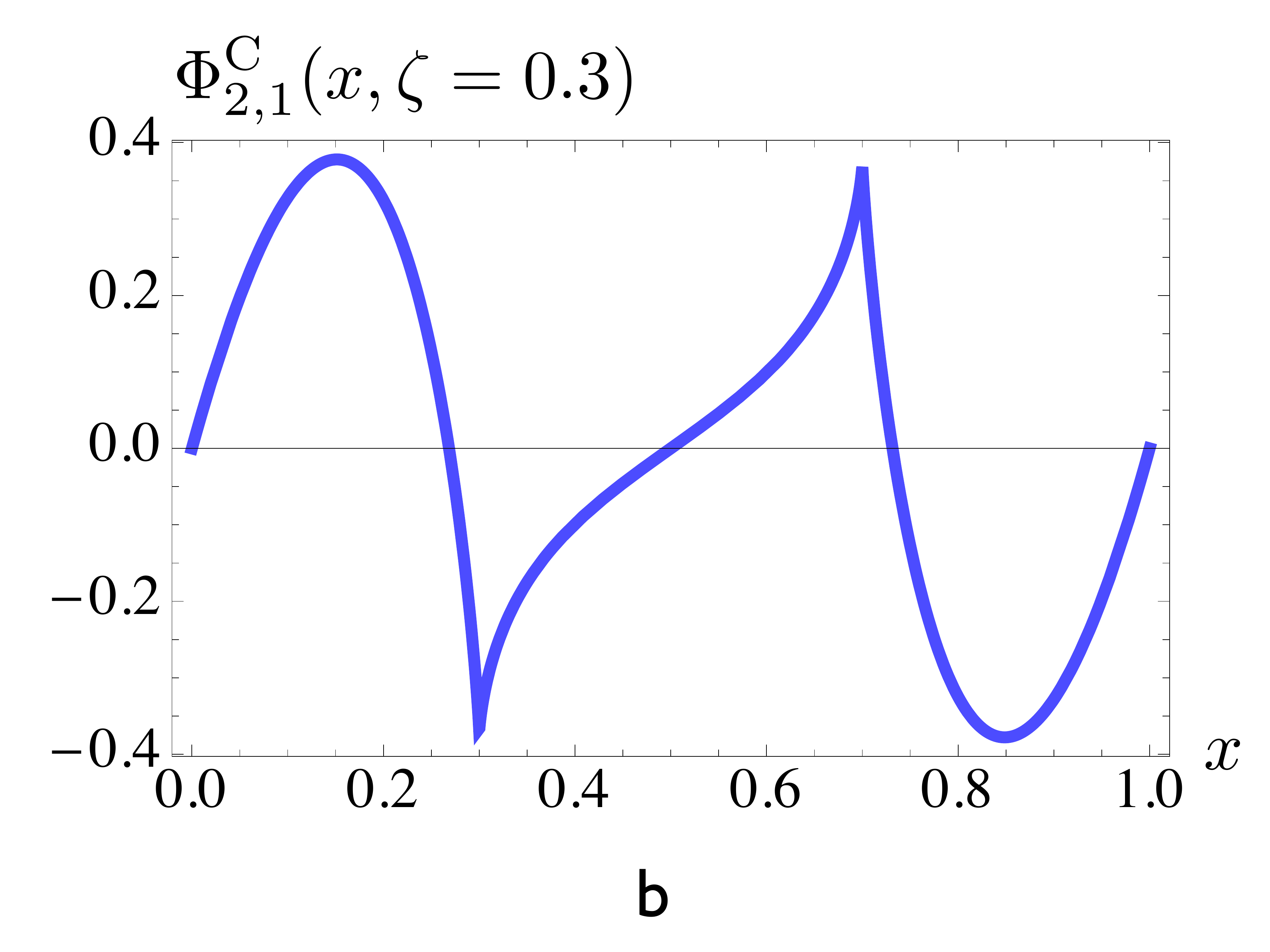}
 \caption{Initial form  $\Phi^\text{C}_{2,0}(x,\zeta=0.2)$  for 
curvy part  (a) and  its first iteration 
(b).}
 \label{phi21l}
  \end{figure}
 Its first iteration  is  given by 
 \begin{align}
   \Phi^\text{C}_{2,1}=
   &
   -\frac{\theta(0<x<\zeta)}{\zeta\bar\zeta}
   \biggl\{
   x[3x-4\zeta+2\zeta^2]
     +2[x\bar{x} (3+x-\zeta)+\bar\zeta
     ]\ln\bar{x}    
  - 2\bar x \bar \zeta \, \ln \bar \zeta 
   +2x\bar x(x-\zeta) \ln x
  \nonumber\\
   &+2x(\zeta-x)\ln(\zeta-x)-2\bar x(\bar\zeta-x)\ln(\bar\zeta-x)
   \biggr\}
   -\{x\rightarrow \bar{x}\}\nonumber\\ &
      -  \frac{ \theta(\zeta<x< \bar \zeta)}{\zeta\bar{\zeta}}
       \biggl\{-(1-2x)\zeta^2-2x(x-\zeta)\ln x 
       +2\bar{x}(\bar{\zeta}-x)\ln\bar x+2(1-2x)\bar{\zeta}\ln\bar{\zeta}\nonumber\\
         &+2x(x-\zeta)\ln(x-\zeta)-2\bar{x}(\bar{\zeta}-x)\ln(\bar\zeta -x)
       \biggr\}
        + v(x) \Phi^\text{L}_{2,0} (x,\zeta) \, .
   \end{align}

  One can see from   Fig. \ref{phi21l}b   that 
evolution spreads  the function  into  to the $\zeta<x<\bar{\zeta}$ interval. 
 Combining the results for the linearized and curvy parts,
 we arrive at the evolution pattern generated   for $\varphi_2 (x,\zeta;t)$
 by the first iteration (see 
Fig.  \ref{phi2evol}a).

 \begin{figure}[h]
 \centering
 \includegraphics[width=7.5cm]{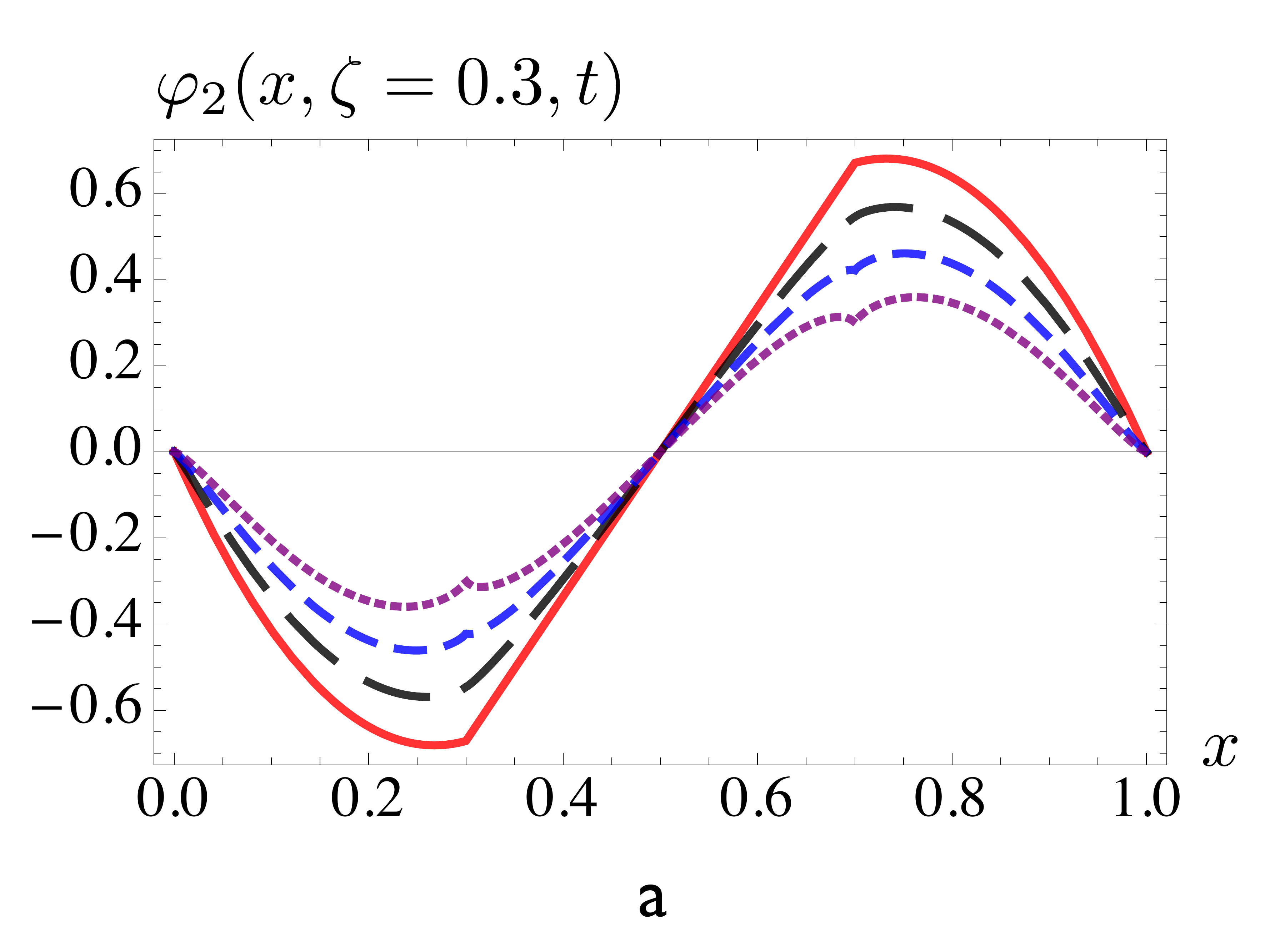}
   \includegraphics[width=7.5cm]{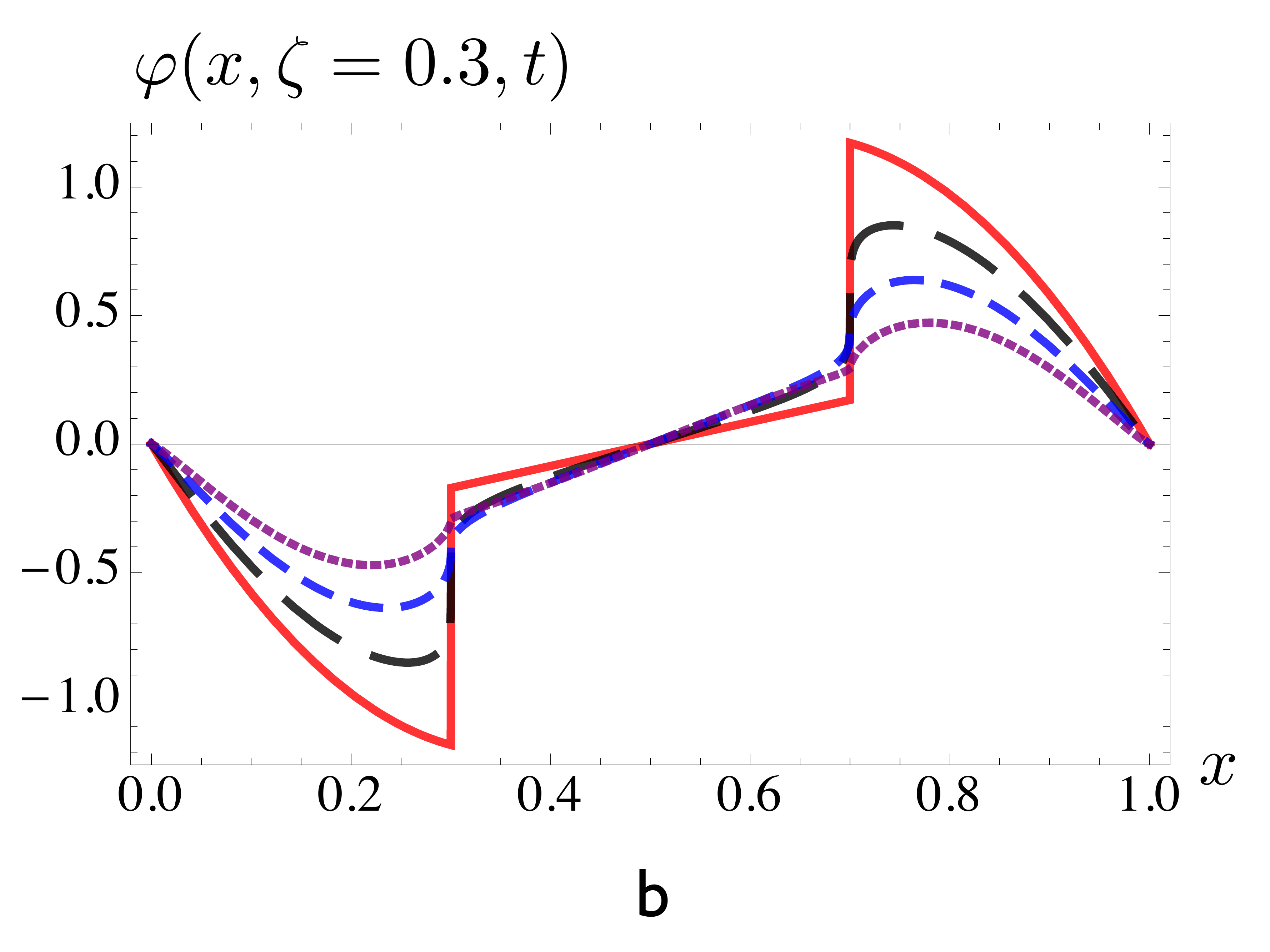}
 \caption{Evolution of   $\phi_{2}(x,\zeta,t)$ at $t=0$, $t=0.2$, $t=0.3$ and $t=0.5$. 
 Evolution of total GDA  $\Phi(x,\zeta,t)$ at $t=0$, $t=0.1$, $t=0.2$ and $t=0.3$}
 \label{phi2evol}
 \end{figure}

Adding  the result for $\varphi_1 (x, \zeta,t)$ obtained in  previous sections, we  
end up with
   the evolution of the  total function 
  $\varphi (x, \zeta,t)$  illustrated  in Fig. \ref{phi2evol}b.

 \end{widetext}

 \section{Summary}
 
In this paper, we described a  new  method for performing  analytic ERBL evolution 
for distribution amplitudes.
Our approach is very efficient in application to    functions that do not vanish at the end points
or have jumps and cusps inside the support region  $0<x<1$.  
Unlike the standard method of expansion  in Gegenbauer polynomials, 
which requires an  infinite number of terms in order to eliminate 
 singularities of initial distributions,  our method  needs 
 only one or two iterations in order to get a reliable 
 and continuous result. 
The method was  illustrated for  two cases of the  initial DA:  for a purely flat DA,
constant in the whole $0\leq x \leq 1$ interval 
and for an antisymmetric DA which  is  constant in each of its two parts
$0 \leq x \leq 1/2$ and $1/2  \leq x \leq 1$.
In  case of  a purely flat DA, the leading term gives $(x \bar x)^t$  evolution
with the change of the evolution parameter $t$.
For the accompanying  factor, two further terms in the $t^N$  expansion 
were found.  In  case of  an antisymmetric flat DA,
there is an extra factor $|1-2x|^{2t}$ that takes care 
of the jump in the middle point $x=1/2$. 
The correction terms   were also calculated.
The results show good convergence for $t \lesssim 1/2$.
It  should  be noted  that for $t \gtrsim 1/2$,  the evolved DA 
is rather close to the asymptotic form, and one can use the standard 
method of the Gegenbauer expansion which is  well  convergent 
for such  functions.
The method was also applied 
for studying the evolution of the (logarithmic $Q^2$ derivative) of the 
 two-photon  GDA. 
  
 The methods developed in the present paper,  may  be extended onto  application 
 to generalized parton distributions. In that case, two strategies are possible.
The first strategy is to  use a direct evolution equation for GPD $H(x,\xi;t)$.
In that case, both the GPD and the evolution kernel depend on the skewness parameter
$\xi$, which is analogous to the parameter $\zeta$ encountered in the two-photon GDA studies.
Another strategy is to use the evolution equation for the double distribution 
$F(\beta,\alpha;t)$. In this case, no skewness parameter is present  in the evolution equation,
and dependence on $\xi$ appears after one performs the conversion of the  double distribution into a GPD.
In both cases, various aspects of our  methods of  analytic evolution may be used.
In particular, GPDs are non-analytic at the border points $x = \pm \xi$, having there cusps,
while model DDs  may have a singular structure (jumps, delta-functions) present in their initial shape.

\section*{Acknowledgements}

This work is supported  by Jefferson Science Associates, LLC under U.S. DOE Contract No.
 DE-AC05-06OR23177.

 \bibliographystyle{apsrev4-1.bst}

\bibliography{GDAevolution_hep_f.bib}

 \end{document}